\documentclass[aps,prl,noshowpacs, twocolumn, nofootinbib,
floatfix,citeautoscript,preprintnumbers]{revtex4-1}

\usepackage[english]{babel}
\usepackage{amssymb}
\usepackage{amsfonts}
\usepackage{amsmath}
\usepackage{eucal}
\usepackage{graphicx}
\usepackage{epsfig}
\usepackage{slashed}

\usepackage[bookmarks=true,colorlinks,linkcolor=blue,urlcolor=blue,citecolor=red]{hyperref}

\newcommand{\be}{\begin{equation}} \newcommand{\ee}{\end{equation}}
\newcommand{\bea}{\begin{eqnarray}} \newcommand{\eea}{\end{eqnarray}}
\newcommand{\beann}{\begin{eqnarray*}}  \newcommand{\eeann}{\end{eqnarray*}}
\newcommand{\bfig}{\begin{figure}} \newcommand{\efig}{\end{figure}}
\newcommand{\ba}{\begin{array}} \newcommand{\ea}{\end{array}}
\newcommand{\bcen}{\begin{center}} \newcommand{\ecen}{\end{center}}
\newcommand{\btab}{\begin{tabular}} \newcommand{\etab}{\end{tabular}}

\newtheorem{Proposition}{Proposition}[section]

\newtheorem{Theorem}{Theorem}[section]
\newtheorem{Lemma}{Lemma}[section]
\newtheorem{Corrolary}{Corrolary}[section]

\newcommand{\bp}{\begin{Proposition}}	\newcommand{\ep}{\end{Proposition}}
\newcommand{\bt}{\begin{Theorem}}	\newcommand{\et}{\end{Theorem}}
\newcommand{\bl}{\begin{Lemma}}		\newcommand{\el}{\end{Lemma}}
\newcommand{\bc}{\begin{Corrolary}}	\newcommand{\ec}{\end{Corrolary}}

\begin{document}

%%%%%%%%%%%%%%%%%%%%%%%%%%%%%%%%%%%%%%
%%%%%%%%%%%%% TITLEPAGE %%%%%%%%%%%%%%
%%%%%%%%%%%%%%%%%%%%%%%%%%%%%%%%%%%%%
\title{Quantum phase transition between a topological and a trivial semimetal \\from holography}

\author{Karl Landsteiner}\email{karl.landsteiner@csic.es}
\author{Yan Liu}\email{yanliu.th@gmail.com}
\author{Ya-Wen Sun}\email{yawen.sun@csic.es}
\affiliation{Instituto de F\'{\i}sica Te\'orica UAM/CSIC, C/ Nicol\'as Cabrera
13-15,\\
Universidad Aut\'onoma de Madrid, Cantoblanco, 28049 Madrid, Spain}

\begin{abstract}
We present a holographic model of a topological Weyl semimetal. A key ingredient is a
time-reversal breaking parameter and a mass deformation. Upon varying the ratio of mass to time-reversal breaking parameter the model undergoes a quantum phase transition from a topologically nontrivial semimetal to a trivial one.
The topological nontrivial semimetal is characterised by the presence of an anomalous Hall effect.  
The results can be interpreted in terms of the holographic renormalization group (RG) flow leading to restoration of time-reversal at the end
point of the RG flow in the trivial phase.

\end{abstract}%
%%%%%%%%%%%%%%%%%%%%%%%%%%%%%%%%%%%%%%
%%%%%%%%%%%% INTRODUCTION %%%%%%%%%%%%
%%%%%%%%%%%%%%%%%%%%%%%%%%%%%%%%%%%%%%
\preprint{IFT-UAM/CSIC-15-124}
\maketitle

Weyl semimetals are an exciting new class of 3D materials with exotic transport properties \cite{wsmreview2,{Hosur:2013kxa}}. They are characterised by pointlike singularities in the Brillouin zone at which conduction and valence bands 
touch. Around these points the electronic quasiparticle excitations can be described by either left- or right-handed Weyl spinors. The Nielsen-Ninomiya theorem guarantees that left- and right-handed Weyl spinors always appear in pairs \cite{Nielsen:1983rb}. When time-reversal symmetry is broken the left- and right-handed
quasiparticles can sit at different points in the Brillouin zone. Effectively the Weyl fermions are separated
by an (axial) vector in the momentum space.
The wave function of a Weyl spinor can be understood as a monopole of the Berry curvature
in momentum space. Left-handed Weyl fermions have monopole charge $+1$ and the right-handed ones have monopole charge $-1$ \cite{Kiritsis:1986re,volovik,{Haldane:2004zz}}. 
Since the monopole charge in momentum
space is a topological invariant it is still present in fermionic two-point correlation functions when interactions are taken into account \cite{Witczak-Krempa:2014nva}. 
However at strong coupling such semiclassical reasoning based on fermionic wave functions or correlators might not be available and
the dynamical variables are physical operators with the quantum numbers of fermion bilinears. 
The question arises then if it is possible to construct a model at strong coupling that has the essential physical properties of a Weyl semimetal, in particular, if there exists any strongly coupled model in which a quantum phase transition between a topological and a topologically trivial state persists even in the absence of the notion of
singularities in the dispersion relations of fermionic two-point correlations functions? A tool to answer these questions 
is the AdS/CFT correspondence (``holography"). It has already proved to be extremely useful for the
understanding of  strongly correlated %transport properties of 
relativistic systems, including superconductors \cite{Hartnoll:2008vx}, strange metals \cite{Liu:2009dm, Cubrovic:2009ye}, lattice systems \cite{Horowitz:2012ky}, etc. In particular, the modern understanding of anomaly related transport phenomena such as the
chiral magnetic and chiral vortical effects, is based to a considerable part on research using holographic models \cite{Erdmenger:2008rm,Banerjee:2008th,Landsteiner:2011iq}\footnote{Previous holographic approaches to the physics of Weyl semimetals \cite{Jacobs:2014nia,Gursoy:2012ie} differ from our approach in that they study holographic fermionic spectral functions.}. 

We first review a quantum field theoretical model model with same local properties
around the band touching points as a Weyl semimetal.  It takes the form of a ``Lorentz breaking'' Dirac
system \cite{Colladay:1998fq} with Lagrangian %(\ref{eq:lagrangian})
\begin{equation}\label{eq:lagrangian}
\mathcal{L} = \bar\Psi \left( i \slashed\partial -  e \slashed{A} - \gamma_5 \vec{\gamma} \cdot \vec{b} + M \right)\Psi\,.
\end{equation}
Here $\slashed X = \gamma^\mu X_\mu$, $A_\mu$ is the electromagnetic gauge potential, $\gamma^\mu$ are the Dirac matrices, and $\gamma_5 = i \gamma_0\gamma_1\gamma_2\gamma_3$
allows us to define left- or right-handed spinors via $(1\pm\gamma_5)\Psi = \Psi_{L,R}$. Without loss of generality we take $\vec{b}= b \hat e_z$. 

\begin{figure}
\begin{center}
 \includegraphics[width=.23\textwidth]{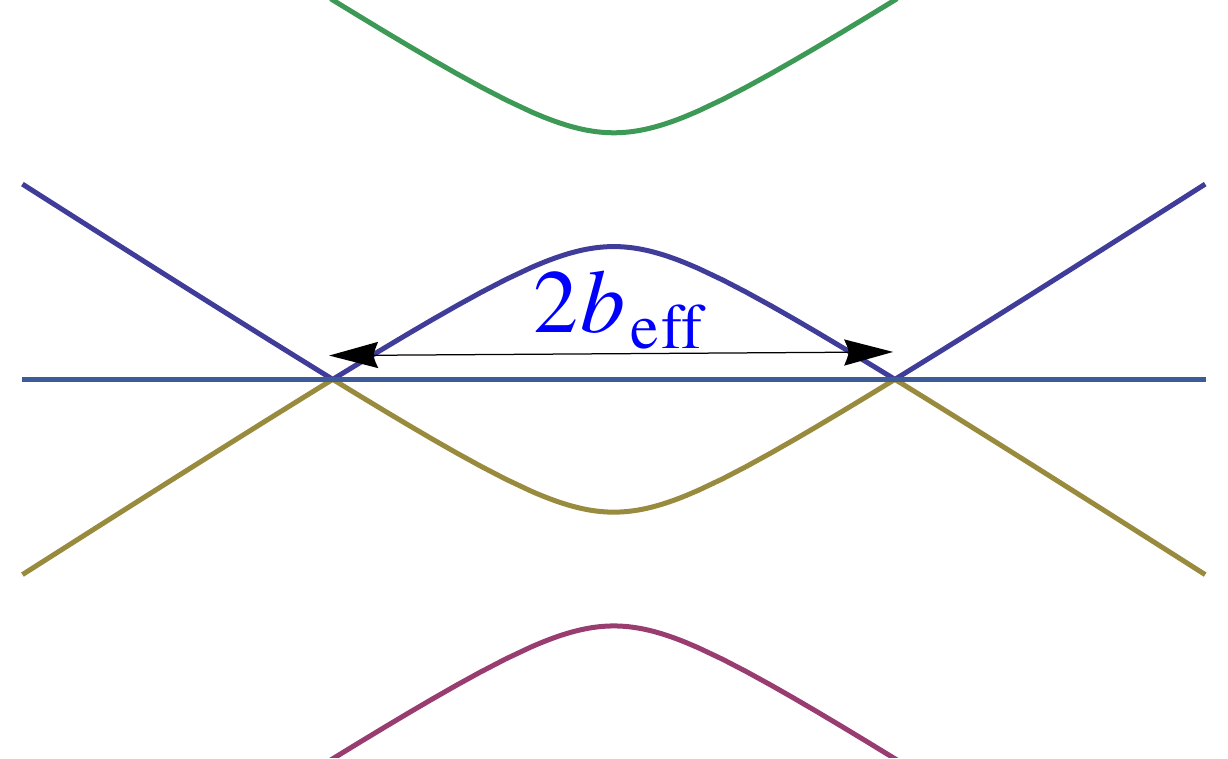}
 \includegraphics[width=.23\textwidth]{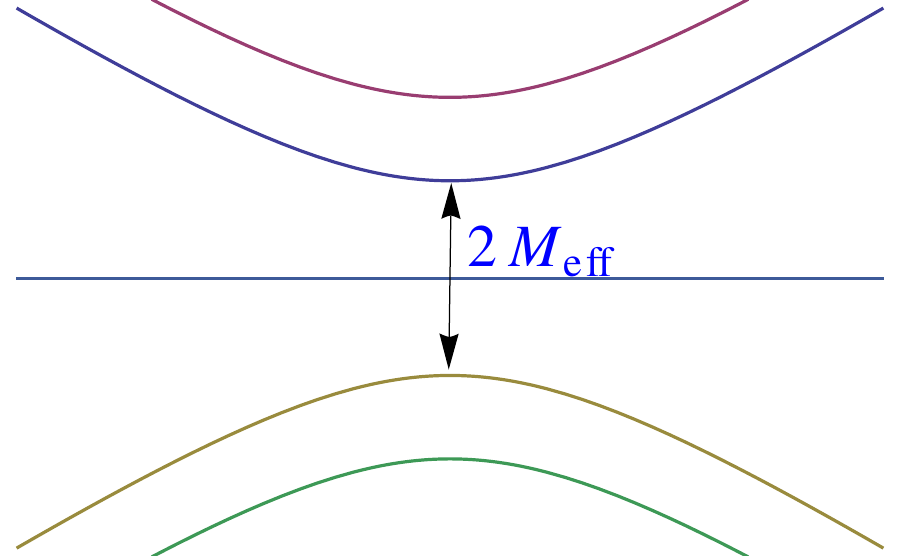}
 \end{center}
 \caption{\label{fig:topphase} \small Left panel: For $b^2>M^2$ there are two Weyl nodes in the spectrum. They are separated by the distance $2\sqrt{b^2-M^2}$ in momentum space.
Right panel: For $b^2<M^2$ the system is gapped with gap $2\sqrt{M^2-b^2}$.}
\end{figure}

The spectrum of (\ref{eq:lagrangian}) is sketched in Fig. \ref{fig:topphase}.
As long as $|b|>|M|$ the spectrum is ungapped. It is characterized by band inversion and at the crossing points the wave function is well-described by Weyl fermion.
The separation of the Weyl cones is given by $2\sqrt{b^2-M^2}$. In this situation the quantum field theoretical model at low energies can be further reduced to 
an effective low energy Lagrangian of the form (\ref{eq:lagrangian}) with $M_\mathrm{eff} =0$ and $\vec b_{\text{eff}} = \sqrt{b^2-M^2} \hat e_z$.
For $|b|<|M|$ the system is gapped and the low energy description is simply one of a massive Dirac fermions
with $b_\mathrm{eff}=0$ and $M_\mathrm{eff} = \sqrt{M^2-b^2}$.
%%%%%%%%%%%%%%%%%%%%%%%%%%%%%%%%%%%%%
%%%%%%%%%%%%%%%%%%%%%%%%%%%%%%%%%%%%%

The axial anomaly
\begin{equation}
 \partial_\mu J^\mu_5 = \frac{1}{16\pi^2} \varepsilon^{\mu\nu\rho\lambda} F_{\mu\nu} F_{\rho\lambda} + 2  M \bar\Psi\gamma_5 \Psi\,
\end{equation}
implies the anomalous Hall effect \cite{Haldane:2004zz,Yang,Xu:2011dn,burkovbalents,Zyuzin:2012tv,Chen:2013mea}
\begin{equation}\label{eq:ahe}
 \vec{J} = \frac{1}{2\pi^2} \vec b_\text{eff} \times \vec E\,.
\end{equation} 
Accordingly the system undergoes a quantum phase transition from the topologically nontrivial Weyl semimetal phase to a trivial insulating phase. This is an example of a topological phase transition not subject to the Landau
classification. In both phases the same symmetries are broken by the nonzero values of the couplings $M,b$.
 Since the topological phase is characterised
by the presence of the Hall effect we can take the Hall conductivity as the order parameter of the transition.
More generally additional massless Dirac fermions might be present. Then the topologically trivial phase would not be gapped but is itself a semimetal. 
The quantum phase transition is then between a topological and a trivial semimetal. This will be the case in our holographic model.

The anomalous Hall effect arises as a one-loop contribution to the polarization tensor.  
Regularization ambiguities \cite{Jackiw:1999qq}  present in the quantum field theoretical model are resolved either
by matching to a tight-binding model \cite{Grushin:2012mt,vazifeh} or by considering anomaly cancellation arising from 
chiral edge states at boundaries (Fermi arcs) \cite{Goswami:2012db}. Essentially this demands a regularization
that is gauge invariant in the presence of axial gauge fields (a source for the axial current).

In the following, we consider the holographic action
\begin{align}\
  S=&\int d^5x \sqrt{-g}\bigg[\frac{1}{2\kappa^2}\Big(R+\frac{12}{L^2}\Big)-\frac{1}{4}F^2-\frac{1}{4}F_5^2\nonumber\\&  
+\frac{\alpha}{3}\epsilon^{\mu\nu\rho\sigma\tau}A_\mu \Big(F^5_{\nu\rho} F^5_{\sigma\tau}+3 F_{\nu\rho} F_{\sigma\tau}\Big)\nonumber\\
&-(D_\mu\Phi)^*(D^\mu\Phi)-V(\Phi)\bigg]\,,\label{eq:holomodel}
\end{align} where $\kappa^2$ is the Newton constant, $L$ is the AdS radius and $\alpha$ is the Chern-Simons coupling constant.\footnote{Note that $\epsilon_{\mu\nu\rho\sigma\beta}=\sqrt{-g}\varepsilon_{\mu\nu\rho\sigma\beta}$ with $\varepsilon_{0123r}=1.$} In holography symmetries of the field theory correspond to gauge fields in AdS space. The electromagnetic $U(1)$ symmetry is represented by the AdS bulk gauge field $V_\mu$ with field strength $F=dV$. The axial $U(1)$ symmetry is represented by the gauge field $A_\mu$ with field strength $F_5 = dA$.\footnote{Note the different conventions from here on.}  It is anomalous and the anomaly is represented in (\ref{eq:holomodel}) by the Chern-Simons part of the action with coupling constant $\alpha$. The gauge invariant regularization corresponds to this choice of Chern-Simons term. It is the unique one that makes the electromagnetic symmetry nonanomalous \cite{Rebhan:2009vc}. 
The mass deformation is introduced via a non-normalizable mode of the scalar field $\Phi$ \cite{Jimenez-Alba:2015awa}. This scalar field is charged only under the axial gauge transformation and its covariant derivative is $D_\mu\Phi = (\partial_\mu - i q A_\mu )\Phi$. The scalar field potential is $m^2 |\Phi|^2 + \frac{\lambda}{2} |\Phi|^4$.
The AdS bulk mass $m^2L^2= -3 $ is chosen such that the dual operator
has dimension three and its source has dimension one. This matches exactly the dimension of the dual of a mass deformation. The electromagnetic and axial currents\footnote{These are the consistent currents. Imposing the equations of motion the vector current $J^\mu$ is conserved whereas the conservation of the the
axial current $J_5^\mu$ is broken by the scalar field and by the anomaly \cite{Jimenez-Alba:2015awa}.
It is also possible to define covariant currents by dropping the Chern-Simons terms \cite{Bardeen:1984pm}. }  are defined as
\begin{align}
\label{eq:consVcur}
J^\mu &= \lim_{r\rightarrow\infty}\sqrt{-g}\Big(F^{\mu r}+4\alpha\epsilon^{r \mu\beta\rho\sigma} A_{\beta} F_{\rho\sigma}  \Big) \,,\\
\label{eq:consAcur}
 J^\mu_5 &= \lim_{r\rightarrow\infty}\sqrt{-g}\Big(F_5^{\mu r}+\frac{4\alpha}{3}\epsilon^{r \mu\beta\rho\sigma} A_{\beta}F^5_{\rho\sigma}  \Big)\,.
\end{align}
The model has been studied before in the probe limit in Ref. \cite{Landsteiner:2015lsa}.

We are looking for solutions that are asymptotically AdS. In addition the holographic analogues of the mass term and the time-reversal breaking parameters in (\ref{eq:lagrangian}) are
introduced via the boundary conditions at $r=\infty$, 
\begin{equation}\label{eq:bcs}
 \lim_{r\rightarrow \infty}\,r\Phi = M~,~~~\lim_{r\rightarrow \infty}A_z = b\,.
\end{equation}
Our ansatz for the zero temperature solution is
\begin{eqnarray}\label{eq:ansatz}
ds^2&=&u(-dt^2+dx^2+dy^2)+\frac{dr^2}{u}+h dz^2\,,\nonumber\\
 A&=&A_z dz\,,~~~\Phi=\phi\,\,.
\end{eqnarray}
Note that due to the conformal symmetry at zero temperature only $M/b$ is a tunable parameter of the system.
In the following we set $2\kappa^2=L=1$.
\vspace{.1cm}\\
\noindent{\bf Critical solution:}
The following Lifshitz-type solution is an exact solution of the system. 
\begin{eqnarray}\label{nh-cs}
ds^2&=&u_0r^2(-dt^2+dx^2+dy^2)+\frac{dr^2}{u_0r^2}+h_0 r^{2\beta}dz^2\,,\nonumber\\
A_z&=&r^\beta,~~~\phi=\phi_0\,.
\end{eqnarray}
 It has the anisotropic Lifshitz-type symmetry $(t,x,y,r^{-1})\to s(t,x,y,r^{-1})$ and $z\to s^\beta z$.
We need to introduce irrelevant deformations to flow it to the UV to match the boundary conditions (\ref{eq:bcs}). 
We can use the scaling symmetry $z\to s z$ to set the coefficient in $A_z$  to be 1. There are four constants $\{u_0, h_0, \beta, \phi_0\}$ that are determined 
by the value of $\lambda$, $m$ and $q$.
We will focus on the simplest case in which there exists only one critical solution and leave the most general $q, \lambda$ analysis for further research. 
To flow this geometry to asymptotic AdS in the UV, we need to consider the following irrelevant perturbation around the Lifshitz fix point
$
u =u_0r^2\big(1+ \delta u\,r^\alpha\big),~~
h= h_0 r^{\beta}\big(1+ \delta h\, r^\alpha\big), ~~
A_z =r^\beta \big(1+ \delta a\, r^\alpha\big),~~
\phi=\phi_0\big(1+ \delta \phi\, r^\alpha\big)
$.
Because of the scaling symmetry, only the sign of $\delta \phi$ is a free parameter and others are fully determined by $\delta \phi=\pm 1$. Numerics shows that only $\delta\phi=-1$ corresponds to asymptotic AdS space at the UV.  
From now we fix the parameters $q=1,\lambda=1/10.$ In this case we have $(u_0, h_0, \beta, \phi_0,\alpha)\simeq (1.468,0.344, 0.407, 0.947,1.315)$ and 
$(\delta u, \delta h, \delta a)\simeq (0.369, -2.797, 0.137)\delta \phi$.  We find the critical value $M/b\simeq 0.744$, which corresponds to the transition point. 
\vspace{.1cm}\\
{\noindent{\bf Topological nontrivial phase:} }
The second kind of solution\footnote{Similar near horizon geometries were found in \cite{Gubser:2009cg,Basu:2009vv} in the context of holographic superconductors.} 
at leading order in the IR is 
\begin{align}\label{nearhor-nt}
u=r^2,~~
h=r^2,~~
A_z=a_1+\frac{\pi a_1^2\phi_1^2}{16 r} e^{-\frac{2 a_1 q}{r}},~~\nonumber\\ \phi=\sqrt{\pi}\phi_1\Big(\frac{a_1 q}{2r}\Big)^{3/2} e^{-\frac{a_1 q}{r}}\,; 
\end{align}
$\lambda$ appears only at higher order terms which become important when $M/b$ is close to the critical value. 
We set $a_1$ to a numerically convenient value and we rescale to $b=1$ later on. 

Starting from the near horizon solution (\ref{nearhor-nt}), we can numerically integrate equations towards the UV and take $\phi_1$ as the shooting parameter to get an 
AdS$_5$ to AdS$_5$ domain wall.  For our chosen values of $\lambda$ and $q$ this kind of solution exists only for $M/b<0.744$. 
\vspace{.1cm}\\
{\noindent{\bf Topological trivial phase:}  }
The third kind of near horizon solution to leading order is  
\begin{equation}\label{nearhor-tt}
u=\big(1+\frac{3}{8\lambda}\big)r^2,~~
h=r^2,~~
A_z=a_1 r^{\beta_1},~~\phi=\sqrt{\frac{3}{\lambda}}+\phi_1 r^{\beta_2}\,,
\end{equation}
where $(\beta_1,\beta_2)=(\sqrt{1+\frac{48q^2}{3+8\lambda}}-1, 2\sqrt{\frac{3+20\lambda}{3+8\lambda}}-2).$
For our choice of $\lambda$ and $q$  $(\beta_1,\beta_2)=(\sqrt{\frac{259}{19}}-1, \frac{10}{\sqrt{19}}-2)$. 
We can set $a_1$ to be 1 and take $\phi_1$ as the shooting parameter to get the AdS$_5$ to AdS$_5$ domain wall.  
We find that this type of solution only exists for $M/b>0.744$.

\begin{figure}
\begin{center}
\includegraphics[width=0.247\textwidth]{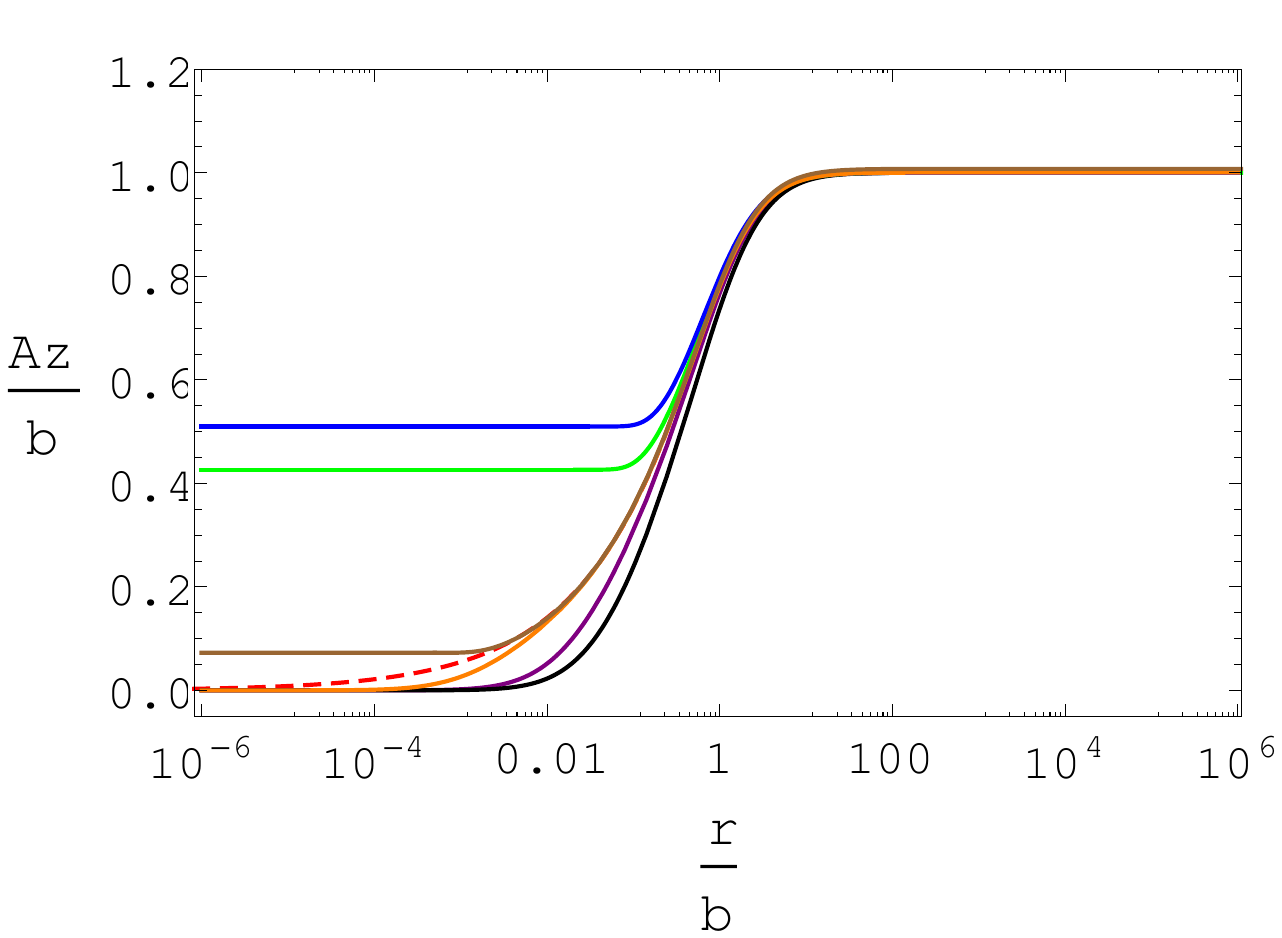}
\includegraphics[width=0.23\textwidth]{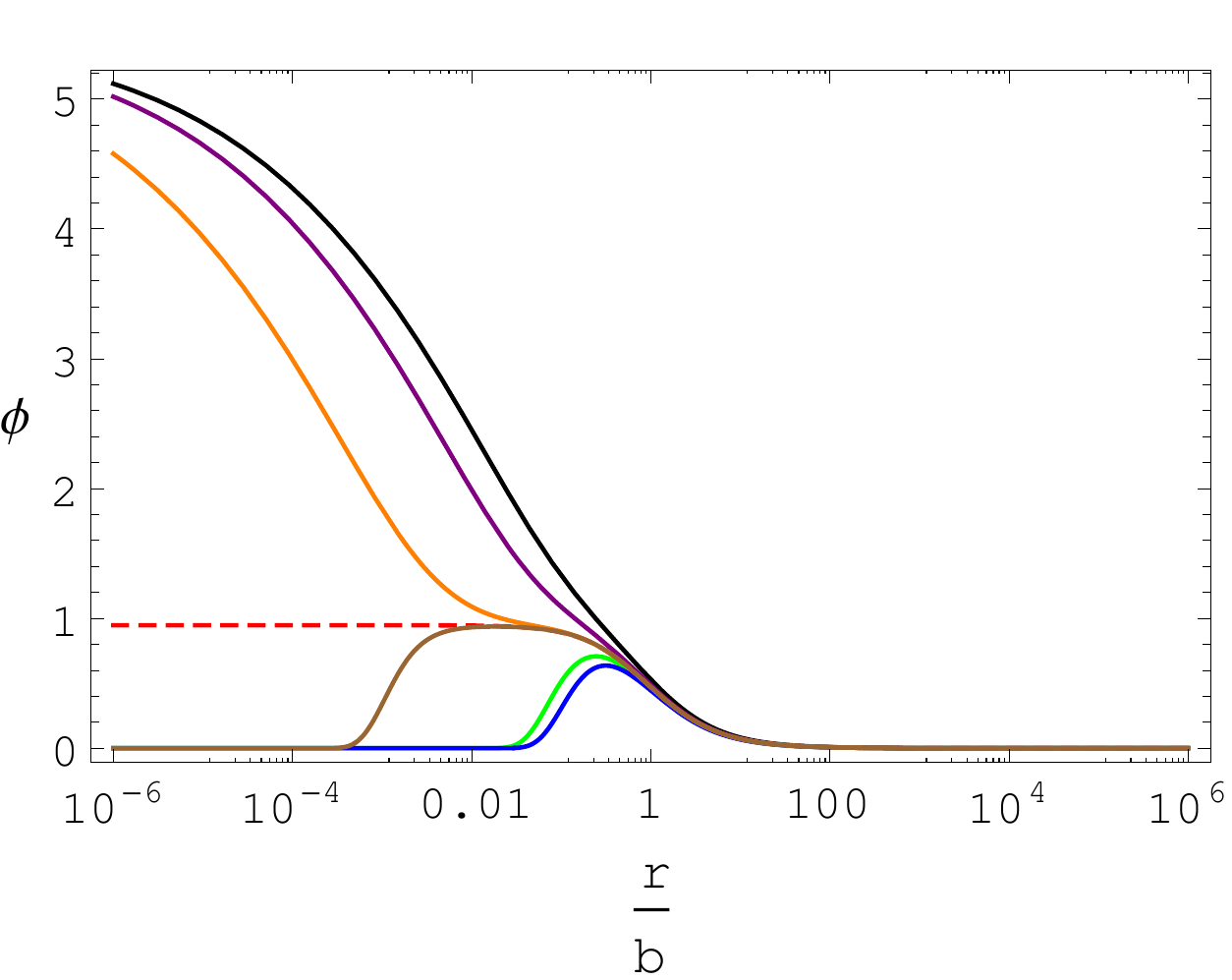}
\end{center}
\caption{\small The bulk profile of background $A_z$ and $\phi$ for $M/b=0.695$ (blue), $0.719$ (green), $0.743$ (brown), 
$0.744$ (red-dashed), $0.745$ (orange), $0.778$ (purple), $0.856$ (black).}
\label{fig:profile}
\end{figure}

In Fig. \ref{fig:profile} we show the behaviour of the scalar field and the gauge field for all three phases at several different values of $M/b$. For a given value of $M/b$ only one of the three types of solutions exists. Note that the value of the gauge field on the horizon matches continuously between the
two phases whereas the value of the scalar field on the horizon jumps discontinuously. Close to the transition point, the near horizon geometry  (\ref{nearhor-nt}) or (\ref{nearhor-tt}) quickly flows to the critical solution in the intermediate IR region as indicated by the brown and orange curves in Fig. \ref{fig:profile}.
Adding standard holographic counterterms we can also compute the free energy density (see the Supplemental Material \cite{sm-ref}).
%Within our numerical accuracy 
We find a %dis
continuous and smooth behaviour at the critical value. 
Note that the free energy is independent of the Chern-Simons coupling and consequently does not probe the topological nature of the transition
in contrast to the Hall conductivity.
%%%%%%%%%%%%%%%%%%%%%%%%%%%%%%%%%%%%%%
\vspace{.1cm}\\
\noindent{\bf Finite temperature solutions:} In order to find finite temperature solutions with a regular horizon at a finite value of $r$ we use the ansatz
\begin{eqnarray}
\label{ansatzforfiniteT}
ds^2&=&-udt^2+\frac{dr^2}{u}+f(dx^2+dy^2)+h dz^2\,,\nonumber\\
A&=&A_z dz\,,~~~\Phi=\phi\,,
\end{eqnarray}
and the conditions that at $r=r_0$ the functions are analytic and that $u$ has simple zero. Using the scaling symmetries of AdS and the constraint from
the equations of motion at the horizon $r=r_0$ we are left with only two dimensionless parameters. In the UV these are mapped to $M/b$ and $T/b$. 
\vspace{.1cm}\\
\noindent{\bf Conductivities} can now be computed with the help of a Kubo formula via retarded correlation functions 
\begin{equation}\label{eq:sigma}
 \sigma_{mn} = \lim_{\omega\rightarrow 0}\frac{1}{i\omega} \langle J_m J_n \rangle (\omega,\vec k =0)\,.
\end{equation}
In holography the retarded Green's functions can be obtained by studying the fluctuations of the gauge fields around the background with infalling boundary conditions at the horizon.

The anomalous Hall conductivity is the off-diagonal part of (\ref{eq:sigma}).
We consider the following fluctuations $\delta V_x=v_x(r) e^{-i\omega t}, ~\delta V_y=v_y(r) e^{-i\omega t}$ and define $v_\pm=v_x+i v_y$,
\begin{equation}\label{eq:flucsxy}
v_\pm''+\bigg(\frac{h'}{2h}+\frac{u'}{u}\bigg)v_\pm'+\frac{\omega^2}{u^2}v_\pm\pm\frac{8\omega\alpha }{u\sqrt{h}}A_z'v_\pm=0\,.
\end{equation}

\begin{figure}
\begin{center}
\includegraphics[width=0.39\textwidth]{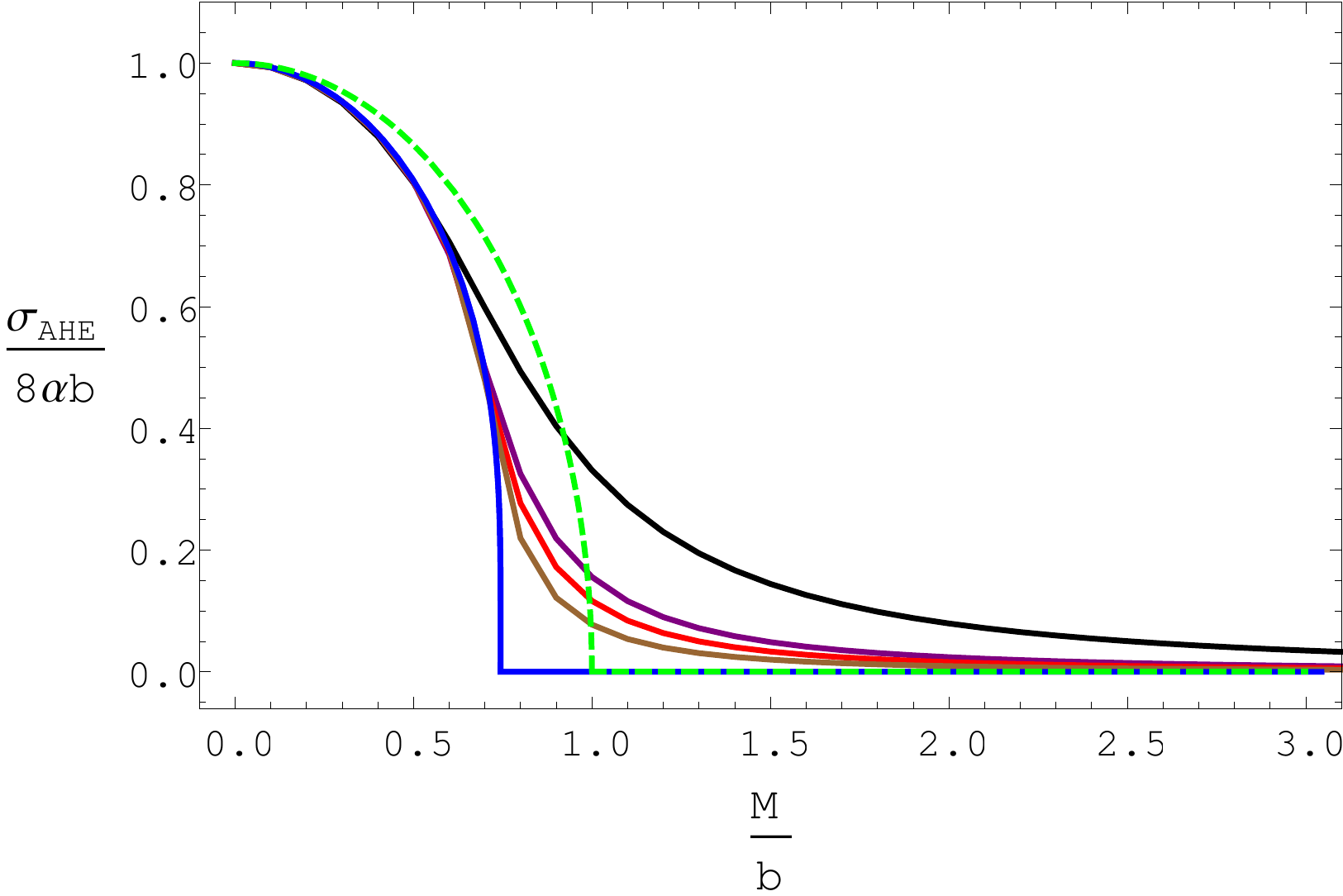}
\end{center}
\caption{\small Anomalous Hall conductivity for different temperatures. The solid lines correspond to our holographic model. For $T=0$ there is a
sharp but continuous phase transition at a critical value of $M/b$ (blue) which becomes a smooth crossover at $T>0$. We show the curves for
$T/b=0.1$ (black), $0.05$ (purple), $0.04$ (red), $0.03$ (brown). For comparision we also show the result for the weak coupling model as a dashed (green) line.
Near the transition the Hall conductivity behaves as $(\sigma_\text{AHE}/b) \propto ((M/b)_c-M/b)^\alpha$ with $\alpha \approx 0.211$ (to be contrasted with the field theory model for which $\alpha=0.5$).}
\label{fig:ahe}
\end{figure}

Note that these are the same for the zero and finite temperature backgrounds.
To solve these equations we follow the usual near-far matching method to first impose ingoing boundary conditions at the near region solutions and match with the far region solutions at a matching region to give the boundary condition for the far region solutions \cite{Faulkner:2009wj}.
To compute the Green's function we normalise the fluctuation to unity at the boundary. The response in the current
is then given by $G_\pm = - u\sqrt{h}  v_\pm'|_{r=\infty} \mp 8 \alpha b \omega$. Note that the second term stems from the Chern-Simons current in (\ref{eq:consVcur}).
We only need to compute the leading order in $\omega$.
For both cases $T=0$ and $T>0$ we can express the result as\footnote{Note that $\sigma_\text{AHE}$ is defined as $\vec{J}=\sigma_\text{AHE} \vec{e}_b\times \vec{E}$ where $\vec{e}_{b}$ is the unit vector along the Weyl nodes separation direction $\vec{b}$. We have $\sigma_\text{AHE}=-\sigma_{xy}=8\alpha A_z(r_0) $. For field theory model, we set $\alpha=\frac{1}{16\pi^2}$ and thus $\sigma_\text{AHE}=\frac{1}{2\pi^2}\sqrt{b^2-M^2}$ in the Weyl semimetal phase. Also note that the consistent currents are used in the main text.}
\begin{equation}
\sigma_{xy}=\frac{G_+-G_-}{2\omega}=-8\alpha A_z(r_0) \,,~~~\sigma_{xx}=\sigma_{yy}= \sqrt{h (r_0)}\,. 
\end{equation}
We emphasise that for $T=$ $r_0=0$ and $h(0)=0$ the diagonal conductivities vanish at zero temperature. 
The anomalous Hall effect (see Fig. \ref{fig:ahe}) is determined by the IR value of the axial gauge field. We can identify $b_\mathrm{eff} = A_z(r=0)$. At zero temperature
it is non vanishing only in the second type of solutions described above. We thus call this the topological nontrivial solution. The third kind
of zero temperature solution is characterised by the restoration of time-reversal invariance at the end point of the holographic RG flow $A_z(0)=0$. 
%%%%%%%%%%%%%%%%%%%%%%%%%%%%%%%%%%%%%
%%%%%%%%%%%%%%%%%%%%%%%%%%%%%%%%%%%%%
\vspace{.1cm}\\
\noindent{\bf Longitudinal conductivity:}
The longitudinal electric conductivity at both finite and zero temperature can be computed from the fluctuation $\delta V_z=v_z e^{-i\omega t}$ with equation of motion 
\begin{equation}
v_z''+\Big(\frac{f'}{f}-\frac{h'}{2h}+\frac{u'}{u}\Big)v_z'+\frac{\omega^2}{u^2}v_z=0\,.
\end{equation}
At zero temperature we substitute $f=u$. 
We again solve it using the semianalytic method of near-far region matching.
At zero temperature we again find $\sigma_{zz} =0$ and for finite temperature we find
\begin{equation}\label{eq:sigmzz}
 \sigma_{zz} = \left.\frac{f}{\sqrt{h}}\right|_{r=r_0}\,.
\end{equation}

%%%%%%%%%%%%%%%%%%%%%%%%%%%%%%%%%%%%%
\begin{figure}
\begin{center}
\includegraphics[width=0.39\textwidth]{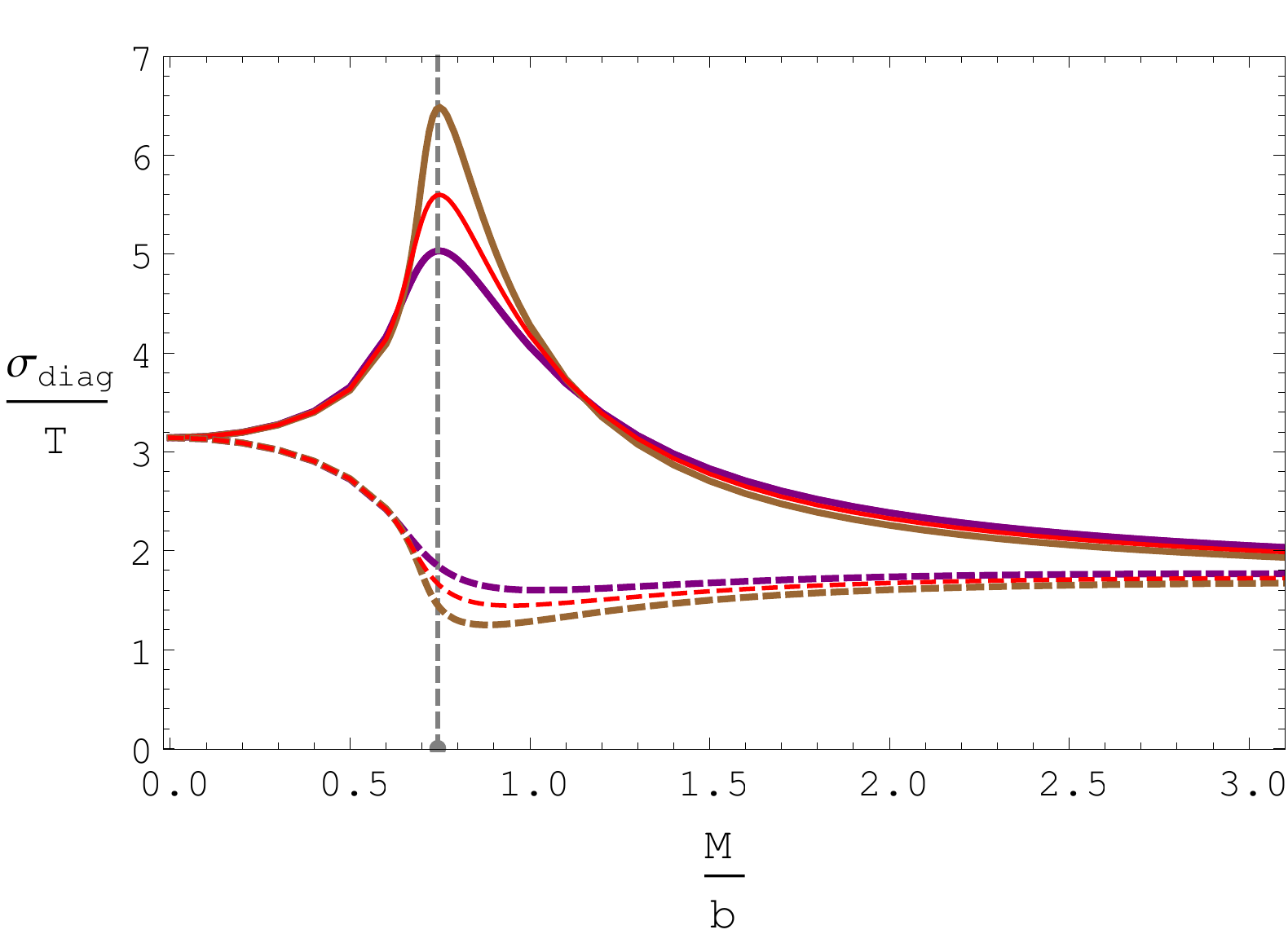}
\end{center}
\caption{\small The transverse and longitudinal electric conductivities for different temperatures. The solid lines are for $\sigma_{xx}=\sigma_{yy}$ and the dashed lines are for $\sigma_{zz}$ from our holographic model with $T/b= 0.05$ (purple), $0.04$ (red), $0.03$ (brown). The dashed gray line is the critical value of $M/b$ at the topological phase transition.
}
\label{fig:con}
\end{figure}
%%%%%%%%%%%%%%%%%%%%%%%%%%%%%%%%%%%%%
%\section{Discussion and outlook}

The three types of background solutions can now be classified according to the anomalous Hall effect. There is a phase for $M/b$ smaller than a critical value in which the axial gauge field flows along the holographic direction towards a constant but nonzero value in the IR. The end point of this holographic flow of the axial gauge
field determines the Hall conductivity $\sigma_\text{AHE}$ (or $\sigma_{xy}$). At $M=0$ the flow is trivial and the Hall response is completely determined by the Chern-Simons current at
the boundary of AdS space. For $M\neq 0$ a nontrivial flow develops, the Hall conductivity has now two parts, a dynamical part, that can only be determined by solving
the equations (\ref{eq:flucsxy}) and the Chern-Simons part determined by the boundary values of the fields. At the critical value (for our choice of parameters this is $(M/b)_\mathrm{c} \simeq 0.744$) the Hall conductivity vanishes. At this value there is a critical
solution with a nontrivial scaling exponent in the $z$-direction. For even larger values of $M/b$ the solution shows no Hall effect. The axial gauge field flows to 
$A_z=0$ in the far IR. In contrast now the scalar field obtains a nontrivial IR value. This corresponds to the effect that the cosmological constant has a different value
in the far IR; i.e., the trivial solution is a domain wall in AdS similar to the zero temperature superconductor solutions described in Ref. \cite{Gubser:2009cg}.  Since in 
holography the cosmological constant is a measure of the effective number of degrees of freedom, we interpret the trivial solution as one in which some of the UV degrees of
freedom are gapped out along the RG flow. We have thus found a holographic zero temperature quantum phase transition between a topological phase characterised
by a non-vanishing Hall conductivity and a topological trivial phase with zero Hall conductivity. All diagonal conductivities vanish at zero temperature.

At $T\neq 0$ the quantum phase transition becomes a smooth crossover behavior. The far IR physics is covered by a horizon at some finite value of  the
holographic coordinate. It is also interesting to observe the behavior of the diagonal conductivities at finite $T$ as a function of $M/b$ (see Fig. \ref{fig:con}). We see that the
transverse diagonal conductivities develop a peak roughly at the critical value whereas the longitudinal one develops a minimum. The height of the peak
and the depth of the minimum grow with temperature. At $M=0$ we simply have $\sigma_{xx,yy,zz}=\pi T$ and for large $M$ the conductivities
tend to a value $\sigma_{xx,yy,zz}= c \pi T$ with $c<1$ and independent of temperature. This is consistent with the interpretation that some but not all degrees of
freedom are gapped out in the trivial phase and that the phase transition is between a topological semimetal and a trivial semimetal. 

We believe our model can serve as a starting point to investigate the paradigm on topological ungapped states of matter in holography and more generally without resorting to the concept of Berry curvature of fermionic wave functions in momentum space.

\subsection*{Acknowledgments}
We thank R. G. Cai,  A. Cortijo, Y. Ferreiro, S. Hartnoll, D. Kharzeev, V. Jacobs, M.A.H. Vozmediano, J. Zaanen for useful discussions.
This work has been supported by project FPA2012-32828 and by the 
Centro de Excelencia Severo Ochoa Programme under grant SEV-2012-0249. 
The work of YWS was also supported by the European Union through a Marie Curie Individual Fellowship MSCA-IF-2014-659135. YL and YWS would like to  thank KITPC for its hospitality and its partial support during the program ``Holographic duality for condensed matter physics".

%%%%%%%%%%%%%%%%%%%%%%%%%%%%%%%%%

%%%%%%%%%%%%%%%%%%%%%%%%%%%%%%%%%
%%%%%%%%%% Merge with supplemental materials %%%%%%%%%%
%\pagebreak
\onecolumngrid
\newpage
%\widetext

\begin{center}
\textbf{\large Supplemental Materials: Quantum phase transition between a topological and a trivial semimetal from holography}
\end{center}
%%%%%%%%%% Merge with supplemental materials %%%%%%%%%%
%%%%%%%%%% Prefix a "S" to all equations, figures, tables and reset the counter %%%%%%%%%%
\setcounter{equation}{0}
\setcounter{figure}{0}
\setcounter{table}{0}
\setcounter{page}{1}
\makeatletter
\renewcommand{\theequation}{S\arabic{equation}}
\renewcommand{\thefigure}{S\arabic{figure}}
\renewcommand{\bibnumfmt}[1]{[S#1]}
\renewcommand{\citenumfont}[1]{S#1}
%%%%%%%%%% Prefix a "S" to all equations, figures, tables and reset the counter %%%%%%%%%%

\section{Finite (and zero) temperature equations and solutions}
At finite temperature, the equations of motion for the ansatz (\ref{ansatzforfiniteT}) are
\bea
u''+\frac{h'}{2h}u'-\bigg(f''+\frac{f'h'}{2h}\bigg)\frac{u}{f}&=&0\,,\nonumber\\
\frac{f''}{f}+\frac{u''}{2u}-\frac{f'^2}{4f^2}+\frac{f'u'}{fu}-\frac{6}{u}+\frac{\phi^2}{2u}\Big(m^2+\frac{\lambda}{2}\phi^2
-\frac{q^2A_z^2}{h}\Big)-\frac{{A_z'}^2}{4h}+\frac{1}{2}\phi'^2&=&0\,,\nonumber\\
\frac{1}{2}{\phi'}^2+\frac{6}{u}-\frac{u'}{2u}\bigg(\frac{f'}{f}+\frac{h'}{2 h}\bigg)
-\frac{f'h'}{2fh}-\frac{f'^2}{4f^2}+\frac{1}{4h}{A_z'}^2
-\frac{\phi^2}{2u}\bigg(m^2+\frac{\lambda}{2}\phi^2+\frac{q^2A_z^2}{h}\bigg)&=&0\,,\nonumber\\
A_z''+\bigg(\frac{f'}{f}-\frac{h'}{2h}+\frac{u'}{u}\bigg)A_z'-\frac{2q^2\phi^2}{u}A_z&=&0\,,\nonumber\\
\phi''+\bigg(\frac{f'}{f}+\frac{h'}{2h}+\frac{u'}{u}\bigg)\phi'-\bigg(\frac{q^2{A_z}^2}{h}+m^2+\lambda\phi^2\bigg)\frac{\phi}{u}&=&0\,.\nonumber
\eea
Now we have three scales $T, b, M$ in the system. 
We have the following three scaling symmetries:\\
(I.)  ~~$(x,y)\to a (x,y)\,,~ f\to a^{-2} f\,$; \\
(II.) ~$z\to a z\,,~ h\to a^{-2} h\,,~ A_z\to a^{-1} A_z\,;$\\
(III.) $r\to a r\,, ~(t,x,y,z)\to (t,x,y,z)/a\,,  ~(u,f, h)\to a^{2} (u, f, h)\,, ~A_z\to a A_z\,.$\\
The scaling symmetry I \& II can be used to set the leading coefficient in front of $r^2$ at the conformal boundary of $f,h$ to be 1. 
The third scaling symmetry can be used to set $r_0$ to be 1 for finite temperature case. 

We start from the near horizon solution and integrate it to the boundary. At the horizon we have 
\bea\label{nhfiniteT1}
u&=&4\pi T(r-r_0)+\dots\,,\\
f&=&f_1-f_1A_{z2}\frac{2\phi_1^2m^2r_0^2-24 r_0^4+\lambda\phi_1^4}{6A_{z1}\phi_1^2 q^2r_0^2}(r-r_0)+\dots\,,\\
h&=&h_1+\dots\,,\\
A_z&=&A_{z1}+A_{z2}(r-r_0)+\dots\,,\\
\label{nhfiniteT2}
 r \phi&=&\phi_1+\frac{\frac{A_{z2}r_0^2}{A_{z1}}\Big(\frac{A_{z1}^2}{h_1}+\frac{m^2}{q^2}\Big)+\phi_1^2\Big(\frac{2}{r_0}+\frac{A_{z2}\lambda}{A_{z1}q^2}\Big)}{2\phi_1}(r-r_0)+\dots
\eea
where $T=\frac{A_{z1}\phi_1^2q^2}{2\pi r_0^2A_{z2}}$. The free parameters at the horizon are $T, r_0, f_1, h_1, A_{z1}, A_{z2}, \phi_1$. With the scaling symmetries, we only have two free parameters. In the dual field theory, these two free  
parameters are $M/b, T/b.$

Near the conformal boundary we have 
\bea\label{as-1}
u&=&r^2-\frac{M^2}{3}+\frac{M^4(2+3 \lambda)}{18}\frac{\ln r}{r^2} -\frac{ M_b}{3r^2}+\dots\,,\\
f&=& r^2-\frac{M^2}{3}+\frac{M^4(2+3 \lambda)}{18}\frac{\ln r}{r^2}+\frac{f_3}{r^2}\dots\,,\\
h&=& r^2-\frac{M^2}{3}+\Big(\frac{M^4(2+3 \lambda)}{18}+\frac{q^2b^2M^2}{2}\Big)\frac{\ln r}{r^2}+\frac{h_3}{r^2}+\dots\,,
\\
A_z&=&b-bM^2q^2\frac{\ln r}{r^2}+\frac{\eta}{r^2}+\dots\,,\\
\label{as-5}
\phi&=&\frac{M}{r}-\frac{\ln r}{6r^3}(2M^3+3b^2Mq^2+3M^3 \lambda)+\frac{O}{r^3}+\dots\,
\eea
with $h_3=\frac{1}{72}(-144f_3+14 M^4-72 MO+9 b^2M^2q^2+9M^4\lambda).$ We have set the non-physical free parameter related to the shift symmetry $r\to r+a$ in the bulk to be zero in the above expansion.  We can integrate the solutions from the horizon with different values of $(M, T, b)$ and finally get the full solutions with different values of $(M/b, T/b)$.

Note that from the first equation of the system we have the radially conserved quantity $(\sqrt{h}(u'f-uf'))'=0$ which further constrain $f_3=-
\frac{1}{3}M_b+\pi Tf_1\sqrt{h_1}.$  This is consistent with the fact that when $T=0$ we have $u=f.$ Since the area of the horizon is with a factor of $f_1\sqrt{h_1}$, we have $f_3=-\frac{1}{3}M_b+\frac{1}{4}Ts$ with $s$ the entropy density of the system in the unit $16\pi G=1$.  Also note that we have another radially conserved quantity $(u'\sqrt{h}f-\frac{h'}{\sqrt{h}}uf-A_zA_z'\frac{uf}{\sqrt{h}})'=0$ from which it follows $2 b \eta-4 MO+b^2M^2q^2-3Ts+4 M_b+(\frac{7}{9}+\frac{\lambda}{2})M^4=0$, thus $h_3=-\frac{1}{3}M_b+\frac{1}{4}Ts-\frac{1}{2}b\eta-\frac{1}{8}b^2M^2q^2.$ These relations are useful when we calculate the free energy of the system.

\begin{figure}[h!]
\begin{center}
\includegraphics[width=0.42\textwidth]{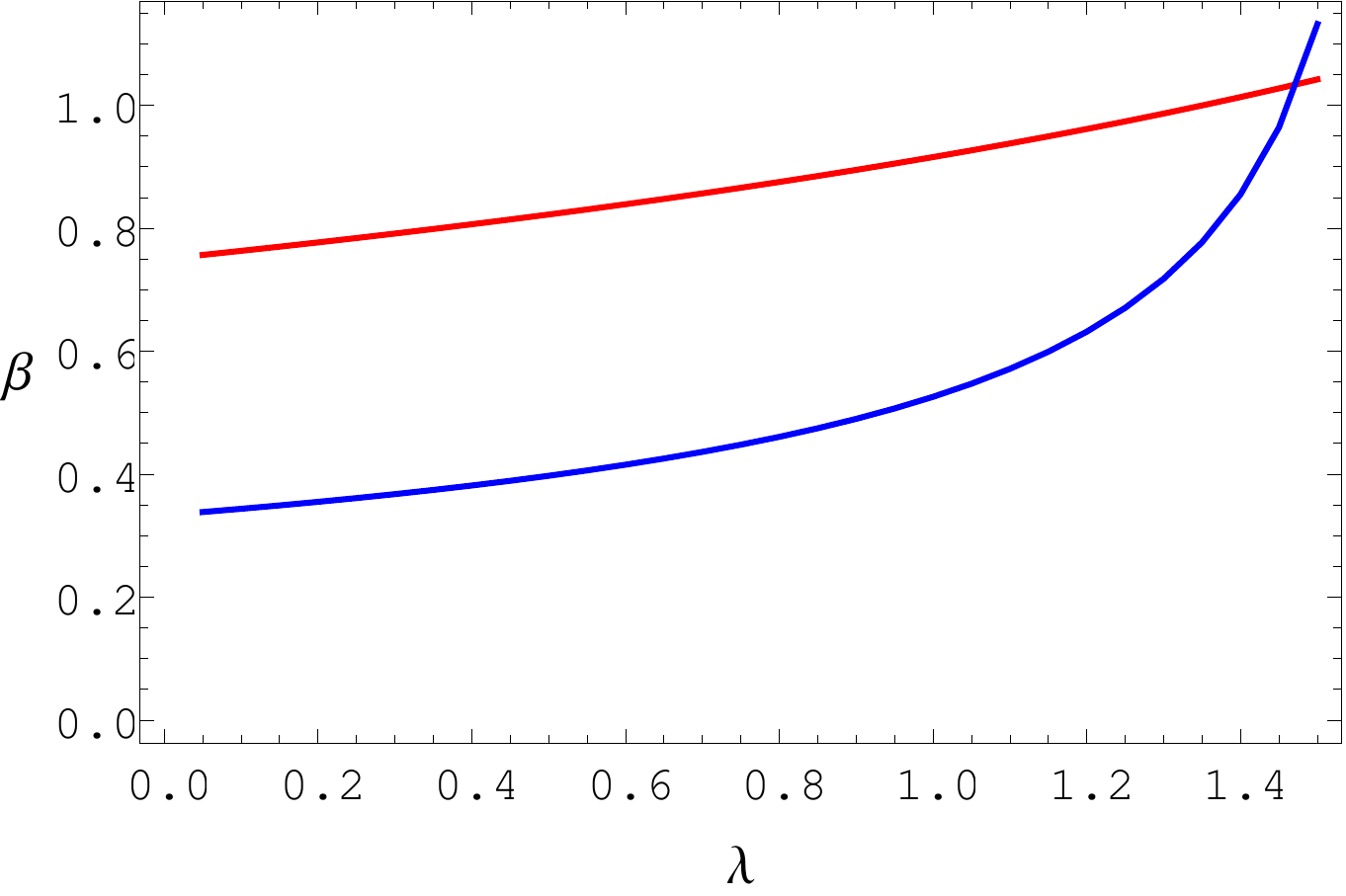}
\hspace{1cm}
\includegraphics[width=0.42\textwidth]{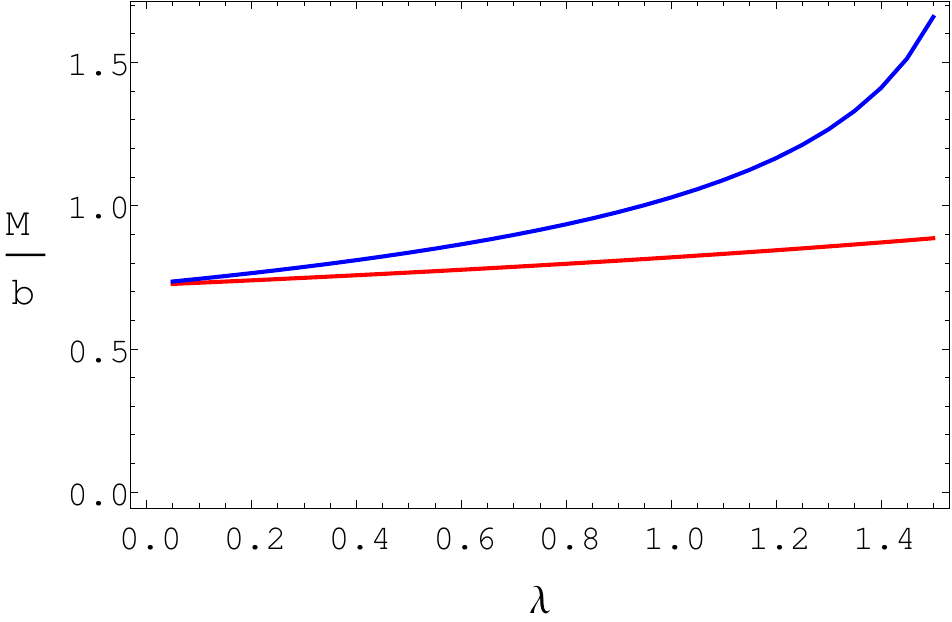}
\end{center}
\caption{\small $q = 1$ (blue) and $q = 3/2$ (red). Left: the anisotropic scaling exponent as a function of $\lambda$. Right: The critical value of $M/b$ as a function of $\lambda$.}
\label{fig:sample}
\end{figure}

At zero temperature, we set $f=u$ in the ansatz (\ref{ansatzforfiniteT}) and this is equivalent to the ansatz (\ref{eq:ansatz}). The equations of motion at zero temperature reduce to four equations.  With the near horizon boundary conditions (\ref{nearhor-nt}-\ref{nearhor-tt}) and after fine-tunning the shooting parameter, we get the whole bulk solutions for the topological nontrivial and trivial phases. With the near horizon condition (\ref{nh-cs}) and the corresponding irrelevant perturbations, we get the zero temperature critical solution.  As an example, the dependance of the anisotropic scaling exponent $\beta$ and the critical value $(M/b)_c$ on $\lambda$ for different $q$ is shown in  Fig. \ref{fig:sample}. 

%%%%%%%%%%%%%%%%%%%%%%%%%%%%%%%%%%%%%%%
\section{Transverse conductivity}
%%%%%%%%%%%%%%%%%%%%%%%%%%%%%%%%%%%%%%%

In this appendix, we calculate the anomalous Hall conductivity in both the topological trivial and nontrivial phases at finite temperature and zero temperature respectively. %\footnote{
Note that in the appendix the results are for covariant current and the consistent current results can be found in the main text. %}. 
The anomalous Hall conductivity ($\sigma_{xy}$) is defined as the retarded Green function of currents $J_x$ and $J_y$, so we consider the following fluctuations in the background of the phases above: $\delta V_x=v_x(r) e^{-i\omega t}, ~\delta V_y=v_y(r) e^{-i\omega t}$. These two modes will not source other fluctuations in the background, so the equations of motion for these two modes are 
\bea
v_x''+\bigg(\frac{h'}{2h}+\frac{u'}{u}\bigg)v_x'+\frac{\omega^2}{u^2}v_x+\frac{8i\omega\alpha }{u\sqrt{h}}A_z'v_y&=&0\,,\\
v_y''+\bigg(\frac{h'}{2h}+\frac{u'}{u}\bigg)v_y'+\frac{\omega^2}{u^2}v_y-\frac{8i\omega\alpha }{u\sqrt{h}}A_z'v_x&=&0\,,
\eea in which $v_x$ and $v_y$ couple together. We can simplify these equations by defining $v_\pm=v_x\pm i v_y$, and now $v_{\pm}$ will not couple to each other
\bea
v_\pm''+\bigg(\frac{h'}{2h}+\frac{u'}{u}\bigg)v_\pm'+\frac{\omega^2}{u^2}v_\pm\pm\frac{8\omega\alpha }{u\sqrt{h}}A_z'v_\pm=0\,.
\eea
We will solve these equations semi-analytically first at finite temperature and then at zero temperature. To solve these equations we follow the usual near far matching method to first impose ingoing boundary conditions at the near region solutions and match with the far region solutions at a matching region to give the boundary condition for the far region solutions.
\vspace{.1cm}\\
%\noindent\comment{Near far matching method}\\
\noindent{\bf Finite temperature}\\
At $T\neq 0$, the background geometry has the near horizon form (\ref{nhfiniteT1} - \ref{nhfiniteT2}) in the near horizon regime $r-r_0 \ll r_0$. We work in the small frequency limit $\omega \ll r_0$ and in the near horizon region, $\omega/(r-r_0)$ can be arbitrarily large thus in the near region $r-r_0\ll r_0$ we have the following form of the equations 
\be\label{finiteTnear}
{v^{(n)}_\pm}''+\frac{1}{r-r_0}{v^{(n)}_\pm}'+\frac{\omega^2}{\big(4\pi T(r-r_0)\big)^2} {v^{(n)}_\pm}=0\,,
\ee
where we kept the leading $\omega^2$ term. Note that this equation also has a linear in $\omega$ term which is subleading compared to other three terms and does not contribute at the leading order. However, for our purpose of calculating the Green function at leading order in $\omega$ we need to keep explicit the dependence of the solutions on $\omega$ at linear order, which means this linear in $\omega$ term has to be treated carefully to give the full dependence of the solutions on $\omega$ at linear order. 

We denote the solution of the equation  (\ref{finiteTnear}) as $v^{(n0)}_\pm= (r-r_0)^{-i\omega/(4\pi T)}.$ Close to the boundary of the near regime, i.e. the matching regime $\omega\ll r-r_0\ll r_0$, the near horizon solution can be expanded as 
\be 
v^{(n0)}_\pm= 1-\frac{i\omega}{4\pi T}\ln(r-r_0)\,.
\ee 
Thus the relative coefficient for the two linearly independent solutions at the matching region $1$ and $\ln(r-r_0)$ is $-\frac{i\omega}{4\pi T}$.

With the solution satisfying ingoing boundary conditions known in the matching region, the easiest way to solve for the linear $\omega$ corrections to the solutions is to work in the matching region. We can denote the solutions as ${v^{(n)}_\pm}={v^{(n0)}_\pm}+\omega v^{(n1)}_{\pm}$ where $v^{(n1)}_{\pm}$ is only sourced by the constant term in $v^{(n0)}_\pm%=0
$ at order $\omega$. After solving for $ v^{(n1)}_{\pm}$ with infalling boundary conditions, it turns out that the full linear order in $\omega$ solution at the matching region is 
\be\label{finiteTbc}  
v^{(n)}_{\pm}=1-i \frac{\omega}{4\pi T} \ln(r-r_0)\mp \frac{8 \alpha \omega}{4\pi T \sqrt{h_1}}A_z(r)\,. 
\ee

This gives the boundary condition at the horizon for the solutions of the far region. 
In the far region $\omega\ll r-r_0$,  we have the equation 
\be
{v^{(f)}_\pm}''+\Big(\frac{h'}{2h}+\frac{u'}{u}\Big){v^{(f)}_\pm}'\pm\frac{8\omega\alpha }{u\sqrt{h}}A_z'{v^{(f)}_\pm}+\frac{\omega^2}{u^2}{v^{(f)}_\pm}=0\,.
\ee 
The solution in the far region can be 
expanded according to $\omega$ and the last term in the equation can be ignored for our purpose. Here we write out the solution up to the first order in $\omega$ after matching with the boundary condition above at the horizon: 
\be
{v^{(f)}_\pm}=c_0+c_1 \int_{r_0+\epsilon}^r \frac{1}{u\sqrt{h}} d\tilde{r}+\omega v^{(f1)}_\pm\,,
\ee 
where $\epsilon$ is a small constant and $c_0=1-i \omega/(4\pi T)\ln \epsilon$, $c_1=-i \omega \sqrt{h_1}$, and $ v^{(f1)}_\pm$ is the order $\omega$ solution sourced by the leading order solution which satisfies the boundary condition of the last term in (\ref{finiteTbc}) at the horizon. $v^{(f1)}_\pm$ can be solved and we have the solution to be \be {v^{(f1)}_\pm}'=\mp \frac{8 \alpha (A_z(r)-A_z(r_0))}{u\sqrt{h}}\,.\ee

%\be{v^{(f)}_\pm}=c_0\pm c_0 \omega \int_r^{\infty}dx \frac{8\alpha}{u\sqrt{h}} \big(A_z-A_z(r_0)\big) .\ee 
%With the boundary condition $c_0 \ln (r-r_0)$ at the horizon, we have solution ${v^{(f)}_\pm}=-c_0 4\pi T\int_r^{\infty}dx \frac{\sqrt{h}|_{r=r_0}}{u\sqrt{h}}.$
%Thus the full solution up to first order in $\omega$ is 
%\be
%v_\pm^{(f)}=
%c_0\pm c_0\omega \int_r^{\infty}dx \frac{8\alpha}{u\sqrt{h}} \big(A_z-A_z(r_0)\big)+ic_0\omega \int_r^{\infty}dx \frac{\sqrt{h}\big{|}_{r=r_0}}{u\sqrt{h}}. 
%\ee

Thus from the solutions we can get $G_\pm=\omega\big(\pm 8\alpha (b-A_z(r_0))+i\sqrt{h(r_0)}\big)$ for covariant current. Since $\sigma_{\pm}=\sigma_{xy}\pm i\sigma_{xx}=\pm \frac{G_\pm}{\omega}$, we have 
\be
\sigma_{xy}=\frac{G_+-G_-}{2\omega}=8\alpha \big(b-A_z(r_0)\big)\,,~~~\sigma_{xx}=\sigma_{yy}= \sqrt{h (r_0)}\,.
\ee

\noindent{\bf Zero temperature}\\
Now we calculate the anomalous Hall conductivity in the zero temperature background with the same technique for both the topological nontrivial and trivial phase. From the explicit form of the near horizon geometry in the main text we can see that for both phases the equations for $v_{\pm}$ are the same in the near region at leading order 
\be
{v^{(n)}_\pm}''+\frac{3}{r}{v^{(n)}_\pm}'+\frac{\omega^2}{r^4} {v^{(n)}_\pm}=0
\ee
and the solution with the infalling boundary ${v^{(n0)}_\pm}= \frac{-i\omega}{r}K_1\big[\frac{-i\omega}{r}\big].$ In the matching regime $\omega\ll r\ll \text{min} \{M, b\}$, the near horizon solution can be expanded as 
\be 
v^{(n0)}_\pm=1-\frac{\omega^2}{4r^2}\big(-1+2\gamma+2\ln\Big[\frac{-i\omega}{2r}\Big]\big)\,.
\ee 
As the two linearly independent solutions at the matching region is $1$ and $1/r^2$, this expansion shows that the infalling solutions corresponds to the solution $1$ at the matching region and the $\omega^2$ term above can be ignored for our purpose.

Similar to the finite temperature case, we also need to calculate the linear order in $\omega$ correction to the near region solution sourced by the infalling leading order solution. At matching region this gives the full linear order in $\omega$ boundary condition as 
\be\label{zeroTbc}
 v^{(n)}_\pm=1+\omega v^{(n1)}_\pm\,,
\ee where 
\be
{v^{(n1)}_\pm}'=\mp\frac{8\alpha (A_z(r)-A_z(0))}{u_0 r^3}\,,
\ee after imposing the infalling boundary condition and this sets the near horizon boundary condition for the far region solution.
%We will take $c_0$ and $-\frac{c_0}{4r^2}\big(-1+2\gamma+2\ln\big[\frac{-i\omega}{2r}\big]\big)$ as the boundary condition at the horizon for the solution of far regime. 
In the far region $\omega\ll r$,  we have the equation 
\be
{v^{(f)}_\pm}''+\Big(\frac{h'}{2h}+\frac{u'}{u}\Big){v^{(f)}_\pm}'\pm\frac{8\omega\alpha }{u\sqrt{h}}A_z'{v^{(f)}_\pm}+\frac{\omega^2}{u^2}{v^{(f)}_\pm}=0\,.
\ee 
The solution in the far region can be 
expanded according to $\omega$ and the last term can be ignored at order $\omega$. Here we write out the solution up to the first order in $\omega$. With the boundary condition (\ref{zeroTbc}) at the horizon, we have the solution $v^{(f)}_\pm=1+ \omega v^{(f1)}_\pm$ where 
\be 
{v^{(f1)}_\pm}' = \mp\frac{8\alpha (A_z(r)-A_z(0))}{u\sqrt{h}}\,.
\ee 

%\int_r^{\infty}dx \frac{8\alpha}{u\sqrt{h}} \big(A_z-A_z(r_0)\big) .$ 

%With the boundary condition $-\frac{c_0}{4r^2}\big(-1+2\gamma+2\ln\big[\frac{-i\omega}{2r}\big]\big)$ at the horizon, we have solution ${v^{(f)}_\pm}=-(-1+2\gamma+2\ln\big[\frac{-i\omega}{2r}\big])\int_r^{\infty}dx c_0\frac{1}{u\sqrt{h}}.$
%Thus the full solution up to first order in $\omega$ is 
%\be
%v_\pm^{(f)}=
%c_0\pm c_0\omega \int_r^{\infty}dx \frac{8\alpha}{u\sqrt{h}} \big(A_z-A_z(r_0)\big)+\mathcal{O}(\omega^2). 
%\ee
With the solutions above, we can get $G_\pm=\omega\big(\pm 8\alpha (b-A_z(0))\big)$ for covariant current. Since $\sigma_{\pm}=\sigma_{xy}\pm i\sigma_{xx}=\pm \frac{G_\pm}{\omega}$, we have 
\be
\sigma_{xy}=\frac{G_+-G_-}{2\omega}=8\alpha \big(b-A_z(0)\big)\,,~~~\sigma_{xx}=\sigma_{yy}= 0\,.
\ee

In the topological trivial phase, we have $\sigma_{xy}=8\alpha b.$ We can see that the formula for the anomalous Hall conductivity is the same for the finite and zero temperature cases. From this formula we can see that the point where $A_z(r_0)$ turns to zero from a finite value signals the topological quantum phase transition.

\section{Longitudinal conductivity}

In this appendix, we calculate the longitudinal electric conductivity at both finite and zero temperature for  the semimetal and the insulator phase at zero density. We consider the fluctuation $\delta V_z=v_z e^{-i\omega t}$ in the background, which does not source other modes at zero density. The equation of motion for $v_z$ is
\be
v_z''+\Big(\frac{f'}{f}-\frac{h'}{2h}+\frac{u'}{u}\Big)v_z'+\frac{\omega^2}{u^2}v_z=0\,.
\ee
We again solve it using the semi-analytic  matching method. 

When $T\neq 0$, in the near horizon regime $r-r_0\ll r_0$ the equation reads 
\be
{v^{(n)}_z}''+\frac{1}{r-r_0}{v^{(n)}_z}'+\frac{\omega^2}{\big(4\pi T(r-r_0)\big)^2} {v^{(n)}_z}=0
\ee
and the infalling solution is $v^{(n)}_z=(r-r_0)^{-i\omega/(4\pi T)}.$ Close to the boundary of the near regime, i.e. the matching regime $\omega\ll r-r_0\ll r_0$, the near horizon solution can be expanded as 
\be v^{(n)}_z=1-\frac{i\omega}{4\pi T}\ln(r-r_0)\,.\ee 
This gives the boundary condition at the horizon for the solution of far region. 
In the far region $\omega\ll r-r_0$,  we have equation ${v^{(f)}_z}''+\big(\frac{f'}{f}-\frac{h'}{2h}+\frac{u'}{u}\big){v^{(f)}_z}'+\frac{\omega^2}{u^2}{v^{(f)}_z}=0$. The solution in the far region can be 
expanded according to $\omega$. Here we write out the solution up to the first order in $\omega$:

 % With the boundary condition $c_0$ at the horizon, we have the solution ${v^{(f)}_z}=c_0+\mathcal{O}(\omega^2).$ 

%With the boundary condition $c_0 \ln (r-r_0)$ at the horizon, we have solution ${v^{(f)}_\pm}=-4\pi T\int_r^{\infty}dx c_0\frac{(f/\sqrt{h})\big{|}_{r=r_0}}{fu/\sqrt{h}}.$
%Thus the full solution up to first order in $\omega$ is 
\be
v_z^{(f)}=
c_0-i\omega \int_{r_0+\epsilon}^{r}d\tilde{r} \frac{(f/\sqrt{h})\big{|}_{r=r_0}}{uf/\sqrt{h}}\,, 
\ee where $c_0=1+O(\omega)$.
%Near the boundary, $v_z^{(f)}\simeq c_0+\frac{i\omega}{2r^2}\frac{f}{\sqrt{h}}\Big{|}_{r=r_0}$. 
Thus the DC longitudinal conductivity at finite temperature is  \be\sigma_{zz}=\frac{f}{\sqrt{h}}\Big{|}_{r=r_0}\, ,\ee which is nonzero and finite.

Let us look at $T=0$. In both the topological nontrivial and trivial phases,  
the near region equation is 
\be
{v^{(n)}_z}''+\frac{3}{r}{v^{(n)}_z}'+\frac{\omega^2}{r^4} {v^{(n)}_z}=0
\ee
and the solution with the infalling boundary ${v^{(n)}_z}= \frac{-i\omega}{r}K_1\big[\frac{-i\omega}{r}\big].$ Close to the boundary of the near regime, i.e. the matching regime $\omega\ll r\ll \text{min}\{M,b\}$, the near horizon solution can be expanded as 
\be v^{(n)}_z= 1-\frac{\omega^2}{4r^2}\big(-1+2\gamma+2\ln\Big[\frac{-i\omega}{2r}\Big]\big)\,.\ee Again this means that the matching region solution $1$ corresponds to the infalling solution.
%We will take $c_0$ and $-\frac{c_0}{4r^2}\big(-1+2\gamma+2\ln\big[\frac{-i\omega}{2r}\big]\big)$ as the boundary condition at the horizon for the solution of far regime. 
In the far region $\omega\ll r$,  we have equation ${v^{(f)}_z}''+\big(\frac{f'}{f}-\frac{h'}{2h}+\frac{u'}{u}\big){v^{(f)}_z}'+\frac{\omega^2}{u^2}{v^{(f)}_z}=0$. The solution in the far region can be 
expanded according to $\omega$. Here we write out the solution up to the first order in $\omega$ to be $v^{(f)}_z=1+\mathcal{O}(\omega^2).$ 
%With the boundary condition $c_0 \ln (r-r_0)$ at the horizon, we have solution ${v^{(f)}_\pm}=-\big(-1+2\gamma+2\ln\big[\frac{-i\omega}{2r}\big]\big)\int_r^{\infty}dx c_0\frac{(f/\sqrt{h})|_{r=r_0}}{f/\sqrt{h}}.$
%Thus the full solution up to first order in $\omega$ is 
%\be
%v_z^{(f)}=c_0+\mathcal{O}(\omega^2). 
%\ee
Thus the DC longitudinal conductivity at zero temperature \be\sigma_{zz}=0\,.\ee
At small frequency, the quantum critical conductivity is linear in $\omega$. This result shows that for both the semimetal and insulator phases, the DC electric conductivity at zero density is zero at $T=0$.

\section{Discussions on the phase transition}

The quantum phase transition between topological state and trivial phase in our letter is quite different from the traditional thermal phase transition, which is formulated in Landau-Ginsburg-Wilson paradigm and is a consequence of symmetry breaking.   Inspired from the recent developments in Weyl semimetal, we identify the topological nontrivial phase with non vanishing anomalous Hall conductivity and topological trivial phase as vanishing anomalous Hall conductivity. In other words, the topological phase transition here is not related to symmetry breaking. Thus the Landau-Ginsburg notion of the phase transition is not valid. Nevertheless, for completeness, in the following we show the free energy of the dual system as a function of $M/b.$

%\subsection{Free energy}
To compute the free energy, we should calculate the regulated on-shell bulk action. We introduce a cutoff surface $r=r_\infty$ and then we add a counterterm to cancel the divergence. The total action is  
\be
S_{\text{ren}}=S+S_{\text{GH}}+S_{\text{c.t.}}
\ee
with the Gibbons-Hawking boundary term $S_\text{GH}=\int_{r=r_\infty}d^4 x \sqrt{-\gamma} (2K)$ and the counterterms
\be
S_{\text{c.t.}}=\int_{r=r_\infty}d^4 x \sqrt{-\gamma}\bigg(-6-|\Phi|^2+\frac{1}{2}(\log r^2)\Big[\frac{1}{4}F^2+\frac{1}{4}F_5^2+|D_m\Phi|^2+(\frac{1}{3}+\frac{\lambda}{2})|\Phi|^4\Big]%+\alpha|\Phi|^4 
\bigg)
\ee
which is necessary to make the variation problem well-defined and the on-shell action finite. 
Here $\gamma_{\mu\nu}$ is the induced metric on the boundary $r=r_\infty$ and the trace of the extrinsic curvature $K=\gamma^{\mu\nu} \nabla_\mu n_\nu$ with $n^\mu$ the outward unit vector normal to the boundary. 
%We have $n_\mu=(0,...0,\frac{1}{\sqrt{u}})$, $\gamma_{\mu\nu}=\text{diag}(-u,f,f,h,0)$ and $K=\frac{\sqrt{u}}{2}\big(\frac{u'}{u}+\frac{2f'}{f}+\frac{h'}{h}\big).$

\begin{figure}[h!]
\begin{center}
\includegraphics[width=0.4\textwidth]{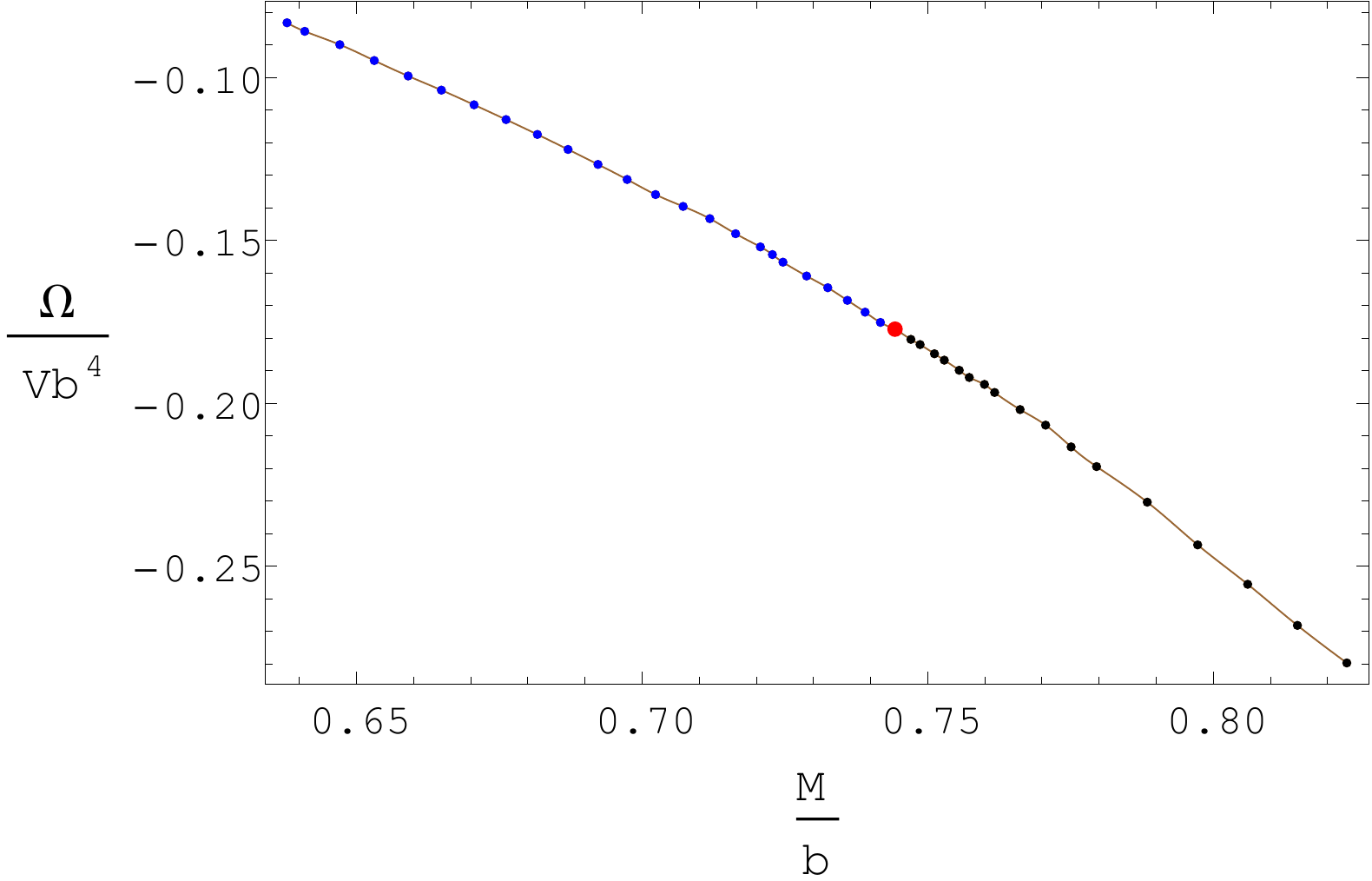}
\end{center}
\caption{\small The free energy density as a function of $M/b$. The red dot is the value of free energy density at the critical point. Close to the transition point from both topological phase and trivial phase, %within our numerical accuracy 
we found the system is continuous and smooth.}
\label{fig:fe}
\end{figure}

%To compute the Euclidean action, we set $t=-i \tau$ with $\tau$ the Euclidean time whose period is the inverse of the temperature. \comment{CHECK}
%Since \bea
%%\sqrt{-g} \mathcal{L}_{\text{on-shell}} &=& \sqrt{-g} \big(2 g^{xx} R_{xx}\big)=-(u\sqrt{h}f')',\nonumber\\
%\sqrt{-\gamma} \mathcal{L}_\text{c.t.}|_{\text{on-shell}}&=&,
%\eea
With the asymptotic data (\ref{as-1}-\ref{as-5}), the on-shell action is $S_{\text{o.s}}=-M_b+Ts+2MO-\frac{7}{36}M^4$
and we have the free energy density $\frac{\Omega}{V}=-\frac{1}{V}S_{\text{o.s}}=M_b-2MO+\frac{7}{36}M^4-Ts.$
%One can make a coordinate transition from (\ref{ansatzforfiniteT}) to the Gaussian normal coordinates to get the dual holographic energy momentum tensor.  
The expectation value of the stress tensor for the dual field theory \cite{Balasubramanian:1999re,
{Myers:1999psa},{deHaro:2000vlm}} can be calculated from  
\be 
T_{ab}= 2(K_{ab}-\gamma_{ab} K)+\frac{2}{\sqrt{-\gamma}}\frac{\delta S_{\text{c.t.}}}{\delta \gamma^{ab}}\,.
\ee
We obtain the resulting total energy density for the dual field theory as $\epsilon=\lim_{r_\infty\to\infty}\sqrt{-\gamma}\langle T^0_{0}\rangle=M_b-2MO+\frac{7}{36}M^4.$ 
%We also have $\langle T^a_{a}\rangle=-4MO-\frac{\lambda}{2}M^4.$  %with $\alpha=-\frac{\lambda}{8}$ we have $\langle T^a_{a}\rangle=-4MO.$
Thus the free energy density is $\frac{\Omega}{V}=\epsilon-Ts.$
The numerical plot for the free energy at zero temperature as a function of $M/b$ is in Fig. \ref{fig:fe}.

%%%%%%%%%%%%%%%%%%%%%%%%%%%%%%%%%

%%%%%%%%%%%%%%%%%%%%%%%%%%%%%%%%%


\begin{thebibliography}{99}

\bibitem{wsmreview2}
O.~Vafek and A.~Vishwanat,
``Dirac Fermions in Solids: From High-Tc Cuprates and Graphene to Topological Insulators and Weyl Semimetals,''
Annual Review of Condensed Matter Physics, 
Vol. 5: 83-112, 
[arXiv:1306.2272 [cond-mat.str-el]].


%\cite{Hosur:2013kxa}
\bibitem{Hosur:2013kxa}
  P.~Hosur and X.~Qi,
  ``Recent developments in transport phenomena in Weyl semimetals,''
  Comptes Rendus Physique {\bf 14}, 857 (2013), 
  [arXiv:1309.4464 [cond-mat.str-el]].
  %%CITATION = ARXIV:1309.4464;%%
  %25 citations counted in INSPIRE as of 24 Apr 2015

\bibitem{Nielsen:1983rb}
  H.~B.~Nielsen and M.~Ninomiya,
  ``Adler-bell-jackiw Anomaly And Weyl Fermions In Crystal,''
  Phys.\ Lett.\ B {\bf 130} (1983) 389.
  %%CITATION = PHLTA,B130,389;%%
  %176 citations counted in INSPIRE as of 24 Apr 2015%

%\cite{Kiritsis:1986re}
\bibitem{Kiritsis:1986re}
  E.~Kiritsis,
  ``A Topological Investigation of the Quantum Adiabatic Phase,''
  Commun.\ Math.\ Phys.\  {\bf 111} (1987) 417.
  %%CITATION = CMPHA,111,417;%%


\bibitem{volovik}
G. E. Volovik, The Universe in a Helium Droplet, (Oxford
University Press, Oxford, 2003).

%\cite{Haldane:2004zz}
\bibitem{Haldane:2004zz} 
  F.~D.~M.~Haldane,
  ``Berry Curvature on the Fermi Surface: Anomalous Hall Effect as a Topological Fermi-Liquid Property,''
  Phys.\ Rev.\ Lett.\  {\bf 93}, 206602 (2004).
 % doi:10.1103/PhysRevLett.93.206602
  %%CITATION = doi:10.1103/PhysRevLett.93.206602;%%


%\cite{Witczak-Krempa:2014nva}
\bibitem{Witczak-Krempa:2014nva}
  W.~Witczak-Krempa, M.~Knap and D.~Abanin,
  ``Interacting Weyl semimetals: characterization via the topological Hamiltonian and its breakdown,''
  Phys.\ Rev.\ Lett.\  {\bf 113} (2014) 136402, 
  [arXiv:1406.0843 [cond-mat.str-el]].
  %%CITATION = ARXIV:1406.0843;%%
  %2 citations counted in INSPIRE as of 20 May 2015


%\cite{Hartnoll:2008vx}
\bibitem{Hartnoll:2008vx} 
  S.~A.~Hartnoll, C.~P.~Herzog and G.~T.~Horowitz,
  ``Building a Holographic Superconductor,''
  Phys.\ Rev.\ Lett.\  {\bf 101}, 031601 (2008)
  [arXiv:0803.3295 [hep-th]].


\bibitem{Liu:2009dm} 
  H.~Liu, J.~McGreevy and D.~Vegh,
  %``Non-Fermi liquids from holography,''
  Phys.\ Rev.\ D {\bf 83}, 065029 (2011)
  [arXiv:0903.2477 [hep-th]].

\bibitem{Cubrovic:2009ye} 
  M.~Cubrovic, J.~Zaanen and K.~Schalm,
  ``String Theory, Quantum Phase Transitions and the Emergent Fermi-Liquid,''
  Science {\bf 325}, 439 (2009) 
  [arXiv:0904.1993 [hep-th]].

\bibitem{Horowitz:2012ky} 
  G.~T.~Horowitz, J.~E.~Santos and D.~Tong,
  ``Optical Conductivity with Holographic Lattices,''
  JHEP {\bf 1207}, 168 (2012)
%  doi:10.1007/JHEP07(2012)168
  [arXiv:1204.0519 [hep-th]].
  %%CITATION = doi:10.1007/JHEP07(2012)168;%%

%\cite{Erdmenger:2008rm}
\bibitem{Erdmenger:2008rm}
  J.~Erdmenger, M.~Haack, M.~Kaminski and A.~Yarom,
  ``Fluid dynamics of R-charged black holes,''
  JHEP {\bf 0901} (2009) 055, 
  [arXiv:0809.2488 [hep-th]].
  %%CITATION = ARXIV:0809.2488;%%
  %210 citations counted in INSPIRE as of 13 May 2015

%\cite{Banerjee:2008th}
\bibitem{Banerjee:2008th}
  N.~Banerjee, J.~Bhattacharya, S.~Bhattacharyya, S.~Dutta, R.~Loganayagam and P.~Surowka,
  ``Hydrodynamics from charged black branes,''
  JHEP {\bf 1101} (2011) 094, 
  [arXiv:0809.2596 [hep-th]].
  %%CITATION = ARXIV:0809.2596;%%
  %199 citations counted in INSPIRE as of 13 May 2015awa

%\cite{Landsteiner:2011iq}
\bibitem{Landsteiner:2011iq}
  K.~Landsteiner, E.~Megias, L.~Melgar and F.~Pena-Benitez,
  ``Holographic Gravitational Anomaly and Chiral Vortical Effect,''
  JHEP {\bf 1109} (2011) 121, 
  [arXiv:1107.0368 [hep-th]].
  %%CITATION = ARXIV:1107.0368;%%
  %67 citations counted in INSPIRE as of 13 May 2015awa


%\cite{Jacobs:2014nia}
\bibitem{Jacobs:2014nia}
  V.~P.~J.~Jacobs, S.~J.~G.~Vandoren and H.~T.~C.~Stoof,
  ``Holographic interaction effects on transport in Dirac semimetals,''
  Phys.\ Rev.\ B {\bf 90} (2014) 045108, 
  [arXiv:1403.3608 [cond-mat.str-el]].
  %%CITATION = ARXIV:1403.3608;%%
  %2 citations counted in INSPIRE as of 18 May 2015
%\cite{Gursoy:2012ie}

%\cite{Gursoy:2012ie}
\bibitem{Gursoy:2012ie}
  U.~Gursoy, V.~Jacobs, E.~Plauschinn, H.~Stoof and S.~Vandoren,
  ``Holographic models for undoped Weyl semimetals,''
  JHEP {\bf 1304} (2013) 127, 
  [arXiv:1209.2593 [hep-th]].
  %%CITATION = ARXIV:1209.2593;%%
  %8 citations counted in INSPIRE as of 18 May 2015

%\cite{Colladay:1998fq}
\bibitem{Colladay:1998fq}
  D.~Colladay and V.~A.~Kostelecky,
  ``Lorentz violating extension of the standard model,''
  Phys.\ Rev.\ D {\bf 58}, 116002 (1998)
  [hep-ph/9809521].
  %%CITATION = HEP-PH/9809521;%%


\bibitem{Yang}
K.~Y.~Yang, Y.~M.~Lu and Y.~Ran,
``Quantum hall effects in a weyl semimetal: Possible application in pyrochlore iridates,''
Phys.\ Rev.\ B {\bf 84} (2011) 075129,
 [arXiv:1105.2353 [cond-mat.str-el]].

%\cite{Xu:2011dn}
\bibitem{Xu:2011dn}
  G.~Xu, H.~Weng, Z.~Wang, X.~Dai and Z.~Fang,
  ``Chern semimetal and Quantized Anomalous Hall Effect in $HgCr_2Se_4$,''
  Phys.\ Rev.\ Lett.\  {\bf 107} (2011) 186806,
  [arXiv:1106.3125 [cond-mat.mes-hall]].
  %%CITATION = ARXIV:1106.3125;%%
  %53 citations counted in INSPIRE as of 24 Apr 2015


\bibitem{burkovbalents}
A.~A.~Burkov and L.~Balents,
``Weyl Semimetal in a Topological Insulator Multilayer," 
 Phys.\ Rev.\ Lett. {\bf 107}, 127205,
 [arXiv:1105.5138 [cond-mat.mes-hall]].


%\cite{Zyuzin:2012tv}
\bibitem{Zyuzin:2012tv}
  A.~A.~Zyuzin and A.~A.~Burkov,
  ``Topological response in Weyl semimetals and the chiral anomaly,''
  Phys.\ Rev.\ B {\bf 86} (2012) 115133, 
  [arXiv:1206.1868 [cond-mat.mes-hall]].
  %%CITATION = ARXIV:1206.1868;%%
  %49 citations counted in INSPIRE as of 13 May 2015

%\cite{Chen:2013mea}
\bibitem{Chen:2013mea}
  Y.~Chen, S.~Wu and A.~A.~Burkov,
  ``Axion response in Weyl semimetals,''
  Phys.\ Rev.\ B {\bf 88}, no. 12, 125105 (2013), 
  [arXiv: 1306.5344 [cond-mat.mes-hall]].
  %%CITATION = 1306.5344;%%
  %29 citations counted in INSPIRE as of 24 Apr 2015



%\cite{Jackiw:1999qq}
\bibitem{Jackiw:1999qq}
  R.~Jackiw,
  á`When radiative corrections are finite but undetermined,''
  Int.\ J.\ Mod.\ Phys.\ B {\bf 14} (2000) 2011, 
  [hep-th/9903044].
  %%CITATION = HEP-TH/9903044;%%
  %84 citations counted in INSPIRE as of 13 May 2015awa


%\cite{Grushin:2012mt}
\bibitem{Grushin:2012mt}
  A.~G.~Grushin,
 ``Consequences of a condensed matter realization of Lorentz violating QED in Weyl semimetals,''
  Phys.\ Rev.\ D {\bf 86} (2012) 045001, 
  [arXiv:1205.3722 [hep-th]].
  %%CITATION = ARXIV:1205.3722;%%
  %33 citations counted in INSPIRE as of 24 Apr 2015

\bibitem{vazifeh}
 M.~M.~Vazifeh and M.~Franz, 
 ``Electromagnetic Response of Weyl Semimetals," 
 Phys.\ Rev.\ Lett. {\bf 111}, 027201 (2013),
 [arXiv:1303.5784 [cond-mat.mes-hall]].


%\cite{Goswami:2012db}
\bibitem{Goswami:2012db}
  P.~Goswami and S.~Tewari,
  ``Axionic field theory of (3+1)-dimensional Weyl semimetals,''
  Phys.\ Rev.\ B {\bf 88} (2013) 245107, 
  [arXiv:1210.6352 [cond-mat.mes-hall]].
  %%CITATION = ARXIV:1210.6352;%%
  %38 citations counted in INSPIRE as of 24 Apr 2015awa



\bibitem{Rebhan:2009vc}
  A.~Rebhan, A.~Schmitt and S.~A.~Stricker,
  ``Anomalies and the chiral magnetic effect in the Sakai-Sugimoto model,''
  JHEP {\bf 1001} (2010) 026
  [arXiv:0909.4782 [hep-th]].
  %%CITATION = ARXIV:0909.4782;%%
  %110 citations counted in INSPIRE as of 24 Apr 2015
%\cite{Gynther:2010ed}
%\bibitem{Gynther:2010ed}
  A.~Gynther, K.~Landsteiner, F.~Pena-Benitez and A.~Rebhan,
  %``Holographic Anomalous Conductivities and the Chiral Magnetic Effect,''
  JHEP {\bf 1102} (2011) 110
  [arXiv:1005.2587 [hep-th]].
  %%CITATION = ARXIV:1005.2587;%%
  %79 citations counted in INSPIRE as of 24 Apr 2015

%\cite{Jimenez-Alba:2015awa}
\bibitem{Jimenez-Alba:2015awa}
  A.~Jimenez-Alba, K.~Landsteiner, Y.~Liu and Y.~W.~Sun,
  ``Anomalous magnetoconductivity and relaxation times in holography,''
  JHEP {\bf 1507} (2015) 117
  [arXiv:1504.06566 [hep-th]].
  %%CITATION = ARXIV:1504.06566;%%
  %5 citations counted in INSPIRE as of 13 Nov 2015


%\cite{Bardeen:1984pm}
\bibitem{Bardeen:1984pm}
  W.~A.~Bardeen and B.~Zumino,
  ``Consistent and Covariant Anomalies in Gauge and Gravitational Theories,''
  Nucl.\ Phys.\ B {\bf 244} (1984) 421.
  %%CITATION = NUPHA,B244,421;%%
  %517 citations counted in INSPIRE as of 24 Apr 2015

%\cite{Landsteiner:2015lsa}
%\cite{Landsteiner:2015lsa}
\bibitem{Landsteiner:2015lsa} 
  K.~Landsteiner and Y.~Liu,
  ``The holographic Weyl semimetal,''
  Phys.\ Lett.\ B {\bf 753}, 453 (2016)
 % doi:10.1016/j.physletb.2015.12.052
  [arXiv:1505.04772 [hep-th]].



%\cite{Gubser:2009cg}
\bibitem{Gubser:2009cg} 
  S.~S.~Gubser and A.~Nellore,
  ``Ground states of holographic superconductors,''
  Phys.\ Rev.\ D {\bf 80}, 105007 (2009)
  [arXiv:0908.1972 [hep-th]].
  %%CITATION = ARXIV:0908.1972;%%

%\cite{Basu:2009vv}
\bibitem{Basu:2009vv} 
  P.~Basu, J.~He, A.~Mukherjee and H.~H.~Shieh,
  ``Hard-gapped Holographic Superconductors,''
  Phys.\ Lett.\ B {\bf 689}, 45 (2010)
  [arXiv:0911.4999 [hep-th]].
  %%CITATION = ARXIV:0911.4999;%%


\bibitem{sm-ref}
See Supplemental Material %\textcolor{blue}{[url]} 
for text, equations and the figure supporting statements made in the Letter, 
which includes Refs. \cite{Balasubramanian:1999re1,{Myers:1999psa1},{deHaro:2000vlm1}}.

%\cite{Balasubramanian:1999re}
\bibitem{Balasubramanian:1999re1} 
  V.~Balasubramanian and P.~Kraus,
  ``A Stress tensor for Anti-de Sitter gravity,''
  Commun.\ Math.\ Phys.\  {\bf 208}, 413 (1999)
  [hep-th/9902121].
  %%CITATION = doi:10.1007/s002200050764;%%


%\cite{Myers:1999psa}
\bibitem{Myers:1999psa1} 
  R.~C.~Myers,
  ``Stress tensors and Casimir energies in the AdS / CFT correspondence,''
  Phys.\ Rev.\ D {\bf 60}, 046002 (1999) 
  [hep-th/9903203].
  %%CITATION = doi:10.1103/PhysRevD.60.046002;%%

%\cite{deHaro:2000vlm}
\bibitem{deHaro:2000vlm1} 
  S.~de Haro, S.~N.~Solodukhin and K.~Skenderis,
  ``Holographic reconstruction of space-time and renormalization in the AdS / CFT correspondence,''
  Commun.\ Math.\ Phys.\  {\bf 217}, 595 (2001)
  %doi:10.1007/s002200100381
  [hep-th/0002230].
  %%CITATION = doi:10.1007/s002200100381;%%
  


%\cite{Faulkner:2009wj}
\bibitem{Faulkner:2009wj} 
  T.~Faulkner, H.~Liu, J.~McGreevy and D.~Vegh,
  ``Emergent quantum criticality, Fermi surfaces, and AdS(2),''
  Phys.\ Rev.\ D {\bf 83}, 125002 (2011)
  [arXiv:0907.2694 [hep-th]].
  %%CITATION = ARXIV:0907.2694;%%


\end{thebibliography}

\begin{thebibliography}{11}

%\cite{Balasubramanian:1999re}
\bibitem{Balasubramanian:1999re} 
  V.~Balasubramanian and P.~Kraus,
  ``A Stress tensor for Anti-de Sitter gravity,''
  Commun.\ Math.\ Phys.\  {\bf 208}, 413 (1999)
  [hep-th/9902121].
  %%CITATION = doi:10.1007/s002200050764;%%


%\cite{Myers:1999psa}
\bibitem{Myers:1999psa} 
  R.~C.~Myers,
  ``Stress tensors and Casimir energies in the AdS / CFT correspondence,''
  Phys.\ Rev.\ D {\bf 60}, 046002 (1999) 
  [hep-th/9903203].
  %%CITATION = doi:10.1103/PhysRevD.60.046002;%%

%\cite{deHaro:2000vlm}
\bibitem{deHaro:2000vlm} 
  S.~de Haro, S.~N.~Solodukhin and K.~Skenderis,
  ``Holographic reconstruction of space-time and renormalization in the AdS / CFT correspondence,''
  Commun.\ Math.\ Phys.\  {\bf 217}, 595 (2001)
  %doi:10.1007/s002200100381
  [hep-th/0002230].
  %%CITATION = doi:10.1007/s002200100381;%%

\end{thebibliography}
\end{document}